\documentclass[12pt,letterpaper]{article} 

 \usepackage{hyperref} 
 \usepackage{graphicx}
 \usepackage{cite}

\begin{document}



\begin{center}
{\LARGE\bf Meson and Baryon dispersion relations\\[2mm] with Brillouin fermions}
\end{center}
\vspace{5pt}

\begin{center}
{\large\bf Stephan D\"urr$\,^{a,b}$}
\,,\,\,
{\large\bf Giannis Koutsou$\,^{c}$}
\,\,and\,\,
{\large\bf Thomas Lippert$\,^{a,b}$}
\\[10pt]
${}^a${\sl Bergische Universit\"at Wuppertal,
Gau{\ss}stra{\ss}e\,20, 42119 Wuppertal, Germany}\\
${}^{b\!}${\sl J\"ulich Supercomputing Center,
Forschungszentrum J\"ulich, 52425 J\"ulich, Germany}\\
${}^c${\sl Cyprus Institute,
CaSToRC, 20 Kavafi Street, Nicosia 2121, Cyprus}
\end{center}
\vspace{5pt}

\begin{abstract}
\noindent
We study the dispersion relations of mesons and baryons built from Brillouin
quarks on one $N_f\!=\!2$ gauge ensemble provided by QCDSF. For quark masses
up to the physical strange quark mass, there is hardly any improvement over the
Wilson discretization, if either action is link-smeared and tree-level clover
improved. For quark masses in the range of the physical charm quark mass, the
Brillouin action still shows a perfect relativistic behavior, while the Wilson
action induces severe cut-off effects.
As an application we determine the masses of the $\Omega_c^0$, $\Omega_{cc}^+$
and $\Omega_{ccc}^{++}$ baryons on that ensemble.
\end{abstract}
\vspace{5pt}



\newcommand{\pad}{\partial}
\newcommand{\hqu}{\hbar}
\newcommand{\ovr}{\over}
\newcommand{\til}{\tilde}
\newcommand{\pri}{^\prime}
\renewcommand{\dag}{^\dagger}
\newcommand{\<}{\langle}
\renewcommand{\>}{\rangle}
\newcommand{\gaf}{\gamma_5}
\newcommand{\nab}{\nabla}
\newcommand{\lap}{\triangle}
\newcommand{\dal}{{\sqcap\!\!\!\!\sqcup}}
\newcommand{\tim}{{\!\times\!}}
\newcommand{\trc}{\mathrm{tr}}
\newcommand{\Trc}{\mathrm{Tr}}
\newcommand{\Mpi}{M_\pi}
\newcommand{\Fpi}{F_\pi}
\newcommand{\Mka}{M_K}
\newcommand{\Fka}{F_K}
\newcommand{\Met}{M_\et}
\newcommand{\Fet}{F_\et}
\newcommand{\Mss}{M_{s\bar{s}}}
\newcommand{\Fss}{F_{s\bar{s}}}
\newcommand{\Mcc}{M_{c\bar{c}}}
\newcommand{\Fcc}{F_{c\bar{c}}}

\newcommand{\gam}{\gamma_\mu}
\newcommand{\gan}{\gamma_\nu}

\newcommand{\al}{\alpha}
\newcommand{\be}{\beta}
\newcommand{\ga}{\gamma}
\newcommand{\de}{\delta}
\newcommand{\ep}{\epsilon}
\newcommand{\ve}{\varepsilon}
\newcommand{\ze}{\zeta}
\newcommand{\et}{\eta}
\renewcommand{\th}{\theta}
\newcommand{\vt}{\vartheta}
\newcommand{\io}{\iota}
\newcommand{\ka}{\kappa}
\newcommand{\la}{\lambda}
\newcommand{\rh}{\rho}
\newcommand{\vr}{\varrho}
\newcommand{\si}{\sigma}
\newcommand{\ta}{\tau}
\newcommand{\ph}{\phi}
\newcommand{\vp}{\varphi}
\newcommand{\ch}{\chi}
\newcommand{\ps}{\psi}
\newcommand{\om}{\omega}

\newcommand{\psb}{\bar{\psi}}
\newcommand{\etb}{\bar{\eta}}
\newcommand{\psh}{\hat{\psi}}
\newcommand{\eth}{\hat{\eta}}
\newcommand{\psd}{\psi^{\dagger}}
\newcommand{\etd}{\eta^{\dagger}}
\newcommand{\qh}{\hat{q}}
\newcommand{\kh}{\hat{k}}

\newcommand{\bdm}{\begin{displaymath}}
\newcommand{\edm}{\end{displaymath}}
\newcommand{\bea}{\begin{eqnarray}}
\newcommand{\eea}{\end{eqnarray}}
\newcommand{\beq}{\begin{equation}}
\newcommand{\eeq}{\end{equation}}

\newcommand{\mr}{\mathrm}
\newcommand{\mb}{\mathbf}
\newcommand{\ri}{\mr{i}}
\newcommand{\Nf}{N_{\!f}}
\newcommand{\Nc}{N_{ c }}
\newcommand{\Nt}{N_{ t }}
\newcommand{\DW}{D_\mr{W}}
\newcommand{\DB}{D_\mr{B}}
\newcommand{\Dst}{D_\mr{st}}
\newcommand{\Dov}{D_\mr{ov}}
\newcommand{\Dke}{D_\mr{ke}}
\newcommand{\Dstm}{D_{\mr{st},m}}
\newcommand{\Dovm}{D_{\mr{ov},m}}
\newcommand{\Dkem}{D_{\mr{ke},m}}
\newcommand{\Dker}{D_{\mr{ke},-\rh/a}}
\newcommand{\MeV}{\,\mr{MeV}}
\newcommand{\GeV}{\,\mr{GeV}}
\newcommand{\fm}{\,\mr{fm}}
\newcommand{\MSbar}{\overline{\mr{MS}}}

\hyphenation{topo-lo-gi-cal simu-la-tion theo-re-ti-cal mini-mum con-tinu-um}


\section{Introduction}


Wilson fermions offer an effective way of regulating the quark sector of QCD.
Their conceptual simplicity entails a one-to-one correspondence between lattice
and continuum flavor, a property which is particularly convenient for studying
flavor physics in the standard model or in one of its extensions.
The main disadvantage is that they induce cut-off effects in physical
observables which are both parametrically and numerically large, i.e.\
$\propto\!a$, where $a$ is the lattice spacing, and the prefactor is sizable
(for a discussion see e.g.\ the recent review \cite{Fodor:2012gf}).

Two technical remedies which have proven useful in mitigating the
discretization effects are clover improvement and link smearing.
The first one changes the parametric behavior to anything between $\al^n a$
and $a^2$ ($\al$ is the strong coupling constant) \cite{Sheikholeslami:1985ij},
depending on whether the coefficient $c_\mr{SW}$ in (\ref{def_Wil}) is adjusted
in perturbation theory or non-perturbatively \cite{Luscher:1996sc}.
The second one concerns the links that enter the Dirac operator, in the
covariant derivative and/or the clover term, and reduces the coefficient that
multiplies the cut-off terms \cite{DeGrand:1998jq,Bernard:1999kc,
Stephenson:1999ns,Zanotti:2001yb,DeGrand:2002vu}.
It turns out that either idea greatly enhances the other's effectiveness
\cite{Capitani:2006ni}; by combining a generic overall link smearing with an
un-sophisticated improvement strategy (e.g.\ the tree-level choice
$c_\mr{SW}=1$) the amount of chiral symmetry breaking becomes as small as
$am_\mr{res}=O(10^{-2})$ \cite{Capitani:2006ni,Durr:2007cy}.

In recent years such tree-level improved fat-link Wilson fermions have proven
extremely successful, in particular in enabling simulations of QCD with
$\Nf=2+1$ dynamical fermions directly at the physical mass point
\cite{Durr:2010aw} (for an overview of the physics results obtained
with such studies see \cite{Fodor:2012gf}).
One can ask whether further action improvements would warrant the potential
increase in CPU time needed to solve the Dirac equation $Dx=b$ for a given
right-hand side $b$.
The goal of this paper is to investigate this question for the case of the
``Brillouin fermion'' proposed in \cite{Durr:2010ch}.
The standard Wilson action (with clover improvement)
\beq
D^\mr{Wil}(x,y)=\sum_\mu \ga_\mu \nab_\mu^\mr{std}(x,y)
-{1\ovr2}I\lap^\mr{\!std}(x,y)+m_0\de_{x,y}
-\frac{c_\mr{SW}}{2}\sum_{\mu<\nu}\si_{\mu\nu}F_{\mu\nu}\de_{x,y}
\label{def_Wil}
\eeq
with $4+m_0=\frac{1}{2\ka}$
uses the simplest possible choice $\nab_\mu^\mr{std}(x,y)=
[U_\mu(x)\de_{x+\hat\mu,y}-U_\mu\dag(x\!-\!\hat\mu)\de_{x-\hat\mu,y}]/2$ of the
symmetric covariant derivative, and the simplest possible choice of the
covariant Laplacian, $\lap^\mr{\!std}$, which is defined with the standard
9-point stencil.
The Brillouin action \cite{Durr:2010ch}
\beq
D^\mr{Bri}(x,y)=\sum_\mu \ga_\mu \nab_\mu^\mr{iso}(x,y)\,
-{1\ovr2}I\lap^\mr{\!bri}(x,y)+m_0\de_{x,y}
-\frac{c_\mr{SW}}{2}\sum_{\mu<\nu}\si_{\mu\nu}F_{\mu\nu}\de_{x,y}
\label{def_Bri}
\eeq
uses the derivative $\nab_\mu^\mr{iso}$ and the Laplacian $\lap^\mr{\!bri}$,
both of which have 81-point stencils chosen to minimize the amount of
rotational symmetry breaking (in the transverse direction in case of
$\nab_\mu^\mr{iso}$, overall in case of $\lap^\mr{\!bri}$).
The precise definition of these discretization schemes is given in the appendix
of \cite{Durr:2010ch}.
Incidentally, it turns out that (\ref{def_Bri}) is an operator which is quite
close to the one of Ref.\cite{Bietenholz:1999km}, in spite of the construction
being based on rather different principles.

The remainder of this article is organized as follows.
In Sec.\,\ref{sec:2} we specify the set of $\Nf=2$ gauge field configurations
that we use to carry out our investigation, and we give further details of the
link smearing and clover improvement (which we use both in the Wilson and in
the Brillouin action, to compare like with like).
Next, Sec.\,\ref{sec:3} contains the precise form of the Wuppertal smearing
that we apply on both the source and the sink side of our propagators, and
describes the procedure by which we tune the mass parameters $\ka$ in
(\ref{def_Wil}, \ref{def_Bri}) to the correct value for the light, strange
and charm quark mass.
Sec.\,\ref{sec:4} contains the central piece of our investigation, a comparison
of the dispersion relation $E(\mb{p})^2$ as a function of the spatial momentum
$\mb{p}^2$ for mesons and baryons built from Wilson and Brillouin fermions.
As a phenomenological application, we compare in Sec.\,\ref{sec:5} the mass of
the $\Omega_c^0$ baryon that we find to experiment, and we give the masses of
the hitherto unobserved states $\Omega_{cc}^+$ and $\Omega_{ccc}^{++}$ on the
ensemble considered.
We summarize our findings in Sec.\,\ref{sec:6}, and arrange details of the
baryon interpolating fields in an appendix.


\section{Ensemble and valence action details \label{sec:2}}


The goal of our investigation is to compare the Wilson (\ref{def_Wil}) and
Brillouin (\ref{def_Bri}) fermion actions in the valence sector, with special
emphasis on the dispersion relation $E^2=E^2(\mb{p}^2)$ for mesons and baryons
composed of such quarks.
We shall use a freely available set of dynamical gauge field configurations,
i.e.\ with the effect of light sea quark loops included.
We select an ensemble out of the $\Nf=2$ collection by QCDSF
\cite{Bietenholz:2010az}, namely the one with
\beq
\be=5.29,\qquad
L/a=40,\qquad
T/a=64,\qquad
\ka_{ud}^\mr{sea}=0.13632
\label{ensemble}
\eeq
and $a\Mpi^\mr{sea}=0.1034(8)$ \cite{Bietenholz:2010az}.
Thus $\Mpi^\mr{sea}L=4.136$, which bears the promise of small finite-size
effects.
QCDSF determines the scale of their $\be=5.29$ ensembles to be
$a^{-1}=2.71(2)(7)$ \cite{QCDSF:2011aa}, tantamount to $a=0.0728(5)(19)\fm$,
which implies $\Mpi^\mr{sea}\simeq280\MeV$ and $L\simeq2.9\fm$.

Either action involves 3-fold APE smeared gauge links, where one smearing
step in $d=4$ space-time dimensions is given by
\beq
V_\mu(x)=P_{SU(3)}
\Big\{
(1\!-\!\al)U_\mu(x)+\frac{\al}{2(d\!-\!1)}\sum_{\pm\nu\neq\mu}
U_\nu(x)U_\mu(x\!+\!\hat\nu)U_\nu\dag(x\!+\!\hat\mu)
\Big\}
\label{def_ape}
\eeq
and we use $\al_\mr{4D}=0.72$.
Here $P_{SU(3)}$ denotes the back-projection to SU(3) as described in
\cite{Durr:2010ch} and $U$ is the unsmeared (original) gauge field.
Note that these four-dimensionally smeared links enter both the (relevant)
covariant derivative and the (irrelevant) clover term of the Wilson and
Brillouin operators (for a summary of the options see \cite{Capitani:2006ni},
but this choice is the simplest and, as far as we can see, the most effective
one).
In addition, the clover coefficient is set to its tree-level value
($c_\mr{SW}=1$) for either action.
In short, this is the same action as used in \cite{Durr:2010ch}, except that
we now use three steps of APE smearing rather than one.


\section{Wuppertal smearing and quark mass tuning \label{sec:3}}


Wuppertal smearing amounts to the spreading of a vector $q(\mb{x},t)$ -- with
non-trivial support on the time-slice $t$ -- within that time-slice by means of
$N_\mr{W}$ operations of the form \cite{Gusken:1989ad}
\beq
q(\mb{x},t)\quad\longrightarrow\quad
\frac{\de_{\mb{x},\mb{y}}+\de_\mr{W}\sum_i\{
V_i    (\mb{x}            ,t)\de_{\mb{x}+\hat{i},\mb{y}}\!+\!
V_i\dag(\mb{x}\!-\!\hat{i},t)\de_{\mb{x}-\hat{i},\mb{y}}\}}
{1+6\de_\mr{W}}\,q(\mb{y},t)
\label{def_wupp}
\eeq
with spreading parameter $\de_\mr{W}$.
The index $i$ runs over the three spatial directions, and an implicit summation
over $\mb{y}$ takes place.
Here the spatial links $V_i(\mb{x},t)$ are to be generated with the
three-dimensional version of the APE smearing (\ref{def_ape}) applied to the
time-slice $t$.
After the Dirac equation $Dp=q$ has been solved with the broadened source $q$,
an identical spreading is applied to the solution $p$ (for each time-slice
separately), i.e.\ we use smeared-smeared propagators with the same smearing on
the source and on the sink side.

In principle Wuppertal smearing has four parameters to adjust, namely
$(N_\mr{3D},\al_\mr{3D})$ of the three-dimensional APE smearing that the
spatial links $V_i(\mb{x},t)$ have undergone, and $(N_\mr{W},\de_\mr{W})$ in
the recipe (\ref{def_wupp}).
For convenience we try to optimize the two pairs separately.

\begin{figure}[!tb]
\includegraphics[width=8.4cm]{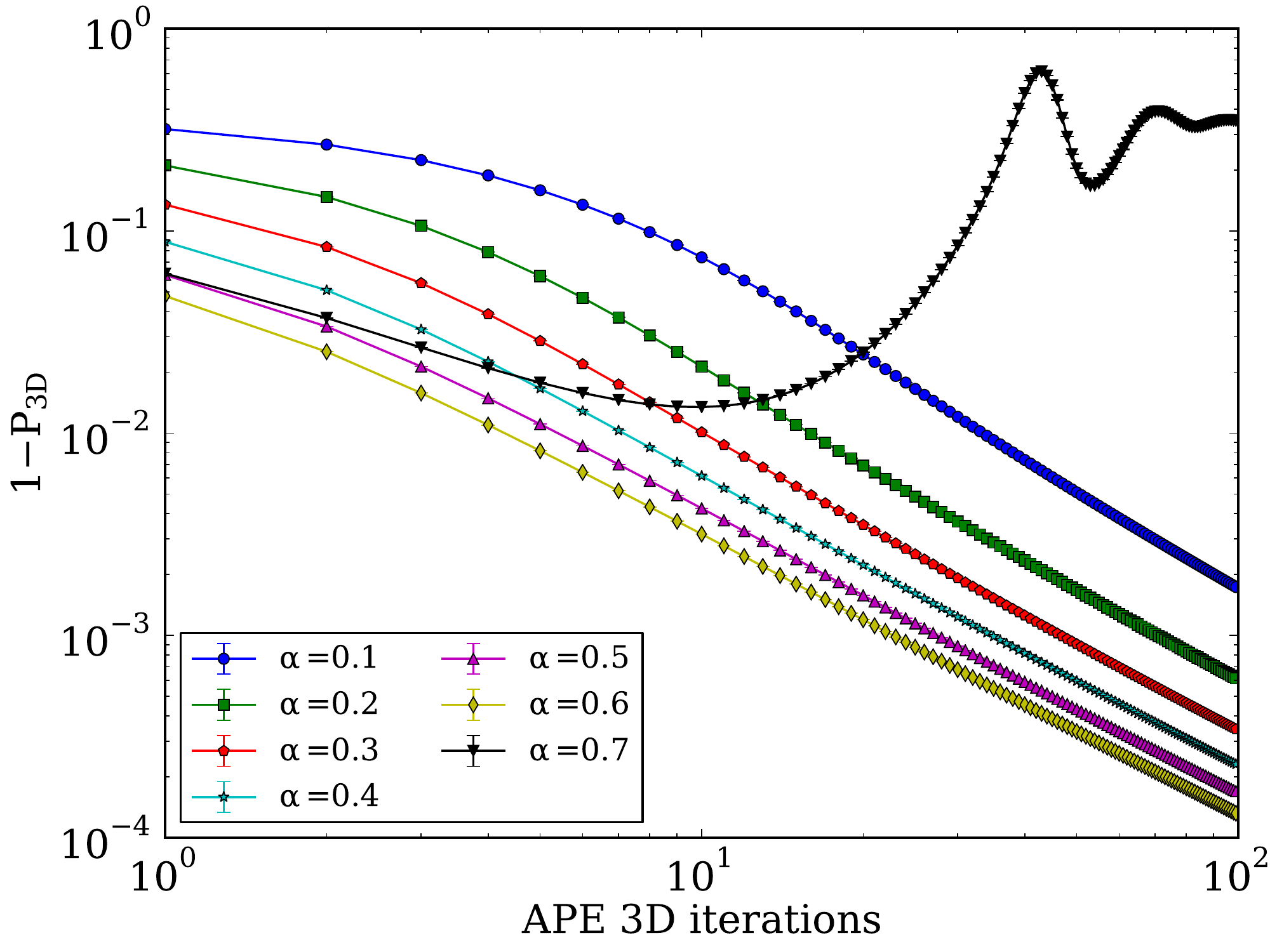}
\includegraphics[width=8.4cm]{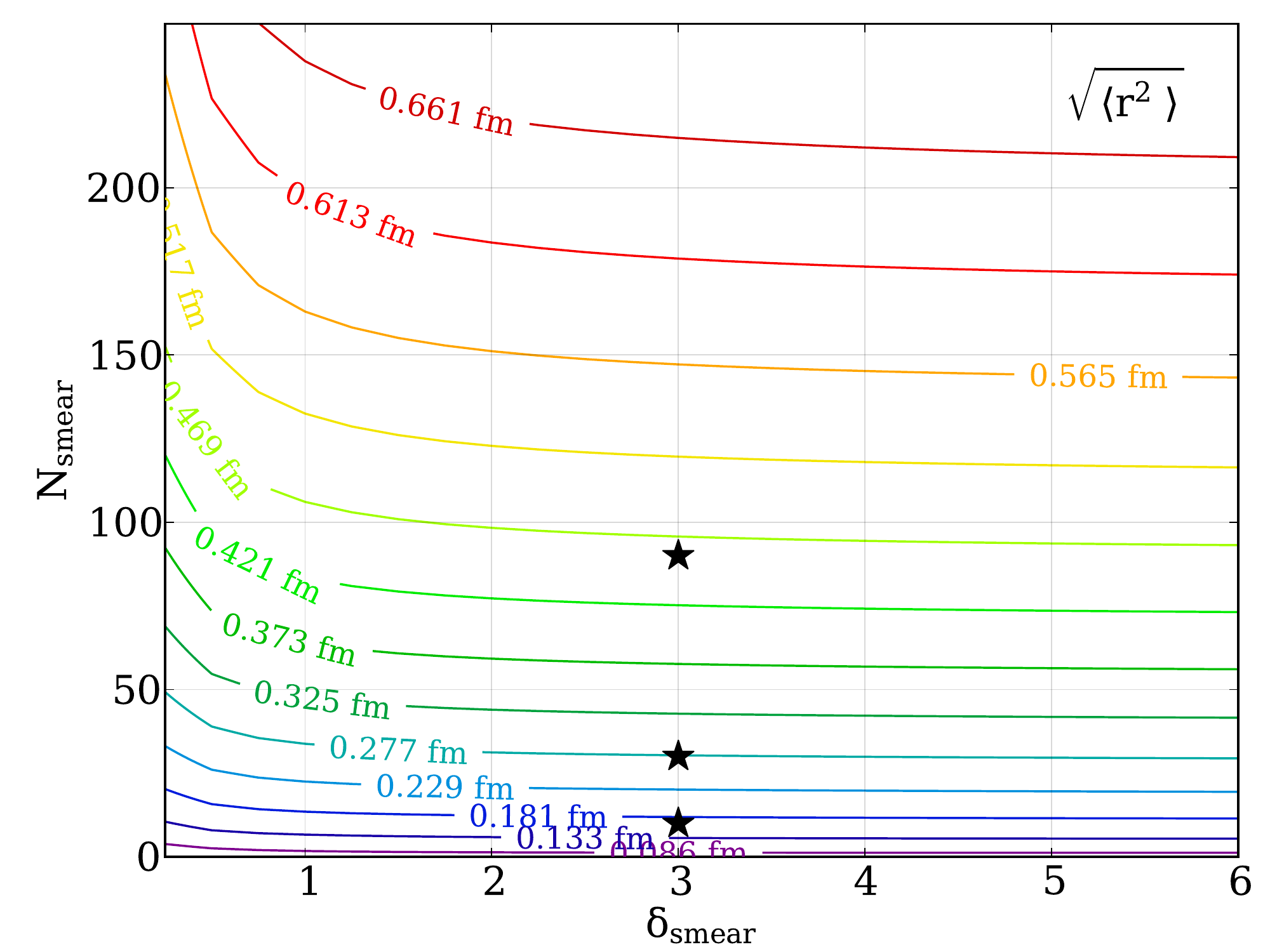}
\caption{\label{fig:ape}\sl
Left: $1-\<\mr{Re}\mr{Tr}(U_\mr{spat})/3\>$ versus the number of 3D APE
smearings for various values of $\al_\mr{3D}$; the perturbative bound is
$\al_\mr{3D}<2/3$ \cite{Capitani:2006ni}. Right: contour plot of
$\<r^2\>^{1/2}$ of a Wuppertal smeared quark source; the asterisks refer to
Fig.\,\ref{fig:wupp}. Either plot is based on 5 configurations.}
\end{figure}

We first study the behavior of the spatial plaquette of the lattices
(\ref{ensemble}) under repeated applications of the three-dimensional version
of the APE recipe (\ref{def_ape}).
Our results are displayed in the left panel of Fig.\,\ref{fig:ape}.
In \cite{Capitani:2006ni} there is the perturbative stability bound
$\al^\mr{APE}<(d\!-\!1)/d$ in $d$ space-time dimensions.
Our results suggest that any $\al_\mr{3D}^\mr{APE}<\al_\mr{3D}^\mr{crit}$
induces, asymptotically, a power-law fall-off of $1-P_\mr{3D}$ and thus
allows us to drive the 3D-plaquette arbitrarily small.
In addition, $\al_\mr{3D}^\mr{crit}$ seems not too far from the perturbative
prediction of $2/3$ \cite{Capitani:2006ni}.
Hence, our recommendation is to apply a large number of 3D APE smearings on
the spatial links that enter the Wuppertal spreading (\ref{def_wupp}), e.g.\
$(N_\mr{3D},\al_\mr{3D})=(300,0.6)$ or $(1000,0.6)$.

\begin{figure}
\includegraphics[width=\textwidth]{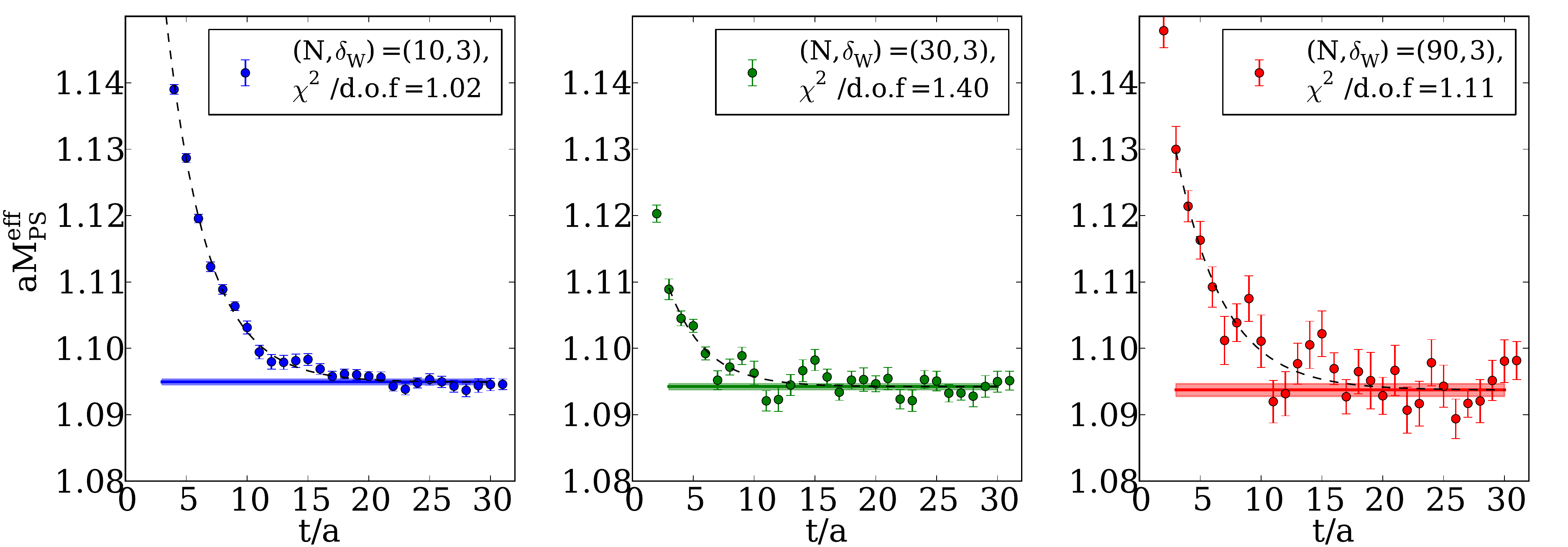}
\caption{\label{fig:wupp}\sl
Effective mass of the pseudoscalar $c\bar{c}$ meson made from Brillouin
fermions with $\ka=0.112429$ for three different widths (cf.\
Fig.\,\ref{fig:ape}) of the Wuppertal smeared sources and sinks.}
\end{figure}

To decide on the second pair $(N_\mr{W},\de_\mr{W})$ in (\ref{def_wupp}) we
first consider the root-mean-square (rms) radius of the smeared source as a
function of these parameters, the former being defined through
\beq
\<r^2\>=\frac
{\sum_\mb{r}\mb{r}^2q\dag(\mb{r})q(\mb{r})}
{\sum_\mb{r}        q\dag(\mb{r})q(\mb{r})}
\;.
\eeq
Results from 5 configurations are shown in the right panel of
Fig.\,\ref{fig:ape}.
As expected, the contours look like hyperbolas, that is the rms radius of $q$
is in the first place a function of the product $N_\mr{W}\de_\mr{W}$.
Next, we consider the effective mass of a meson made from $\ka=0.112429$
Brillouin fermions (which is in the vicinity of $\ka_c^\mr{Bri}$, see below)
for $(N_\mr{W}=10,30,90,\de_\mr{W}=3)$, with results presented in
Fig.\,\ref{fig:wupp}.
In this range, a higher iteration count appears not to decrease the coupling to
excited states, but rather seems to induce more noise in the correlators.
As a consequence we decide to stay with the conservative parameter set
$(N_\mr{W}=10,\de_\mr{W}=0.5)$.

To compare like with like, we wish to compare the dispersion relation for
mesons and baryons put together from Wilson and Brillouin fermions at a fixed
value of the light, strange or charm quark mass.
This is most conveniently done by first tuning the two $\ka$-values to get
common values of $a\Mpi$, $a\Mss$ and $a\Mcc$, respectively, for either
discretization.
For the mesons we use the $PP$ correlators, and given their cosh-form it is
advantageous to define the effective mass as
\beq
aM_\mr{eff}(t)=\frac{1}{2}
\log\bigg(\frac
{C(t\!-\!1)+\sqrt{C(t\!-\!1)^2-C(T/2)^2}}
{C(t\!+\!1)+\sqrt{C(t\!+\!1)^2-C(T/2)^2}}
\bigg)
\eeq
because this modification remedies the fall-off that the effective mass would
show near the center of the box if the generic definition
\beq
aM_\mr{eff}(t)=\frac{1}{2}
\log\bigg(\frac
{C(t\!-\!1)}
{C(t\!+\!1)}
\bigg)
\eeq
would be used.
For baryons with (exact or approximate) projection to a definite parity (see
Sec.\,\ref{sec:4} below) we use the latter form, since these correlators do not
show the cosh form.

\begin{figure}
\includegraphics[width=8.4cm]{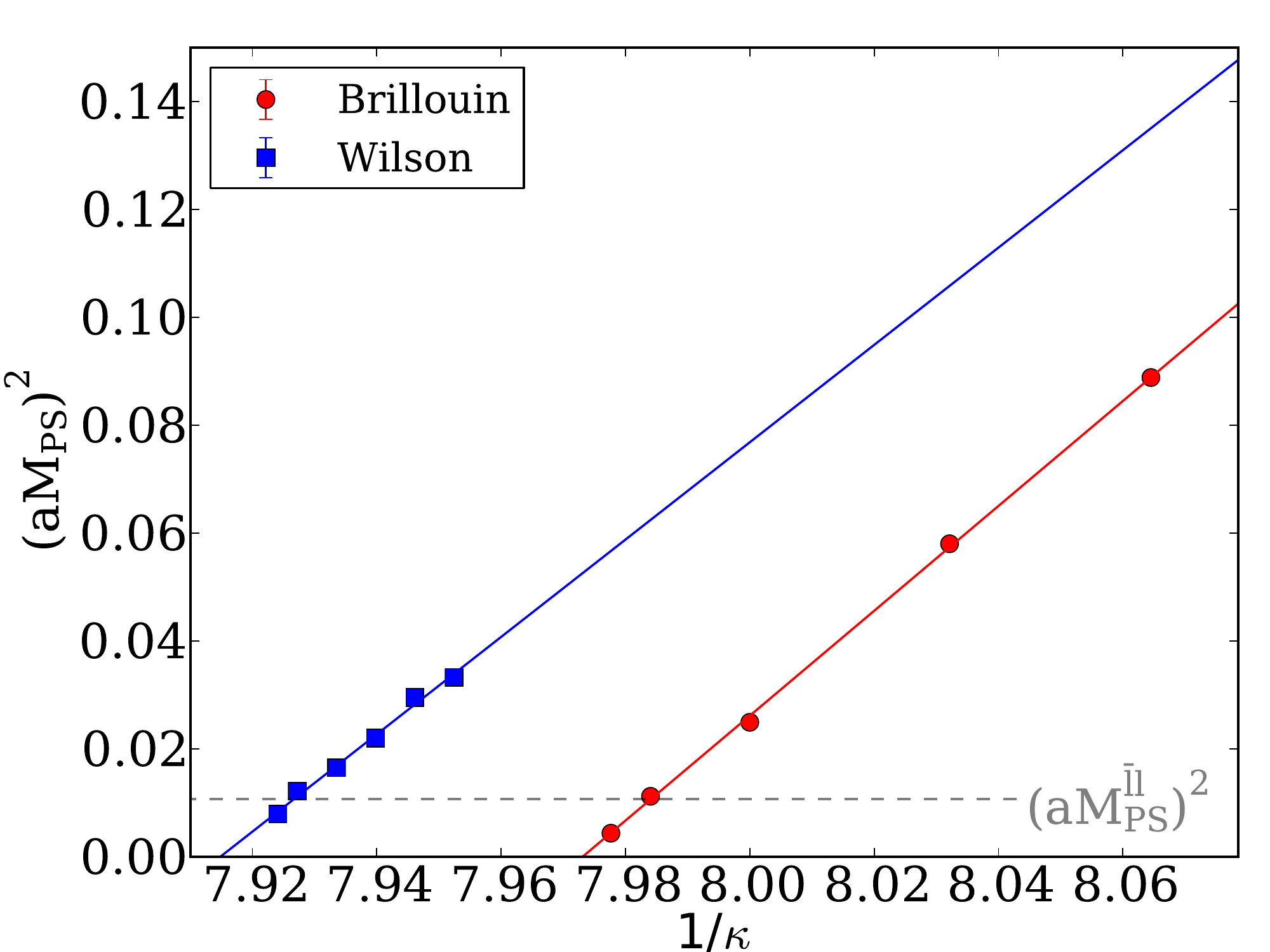}
\includegraphics[width=8.4cm]{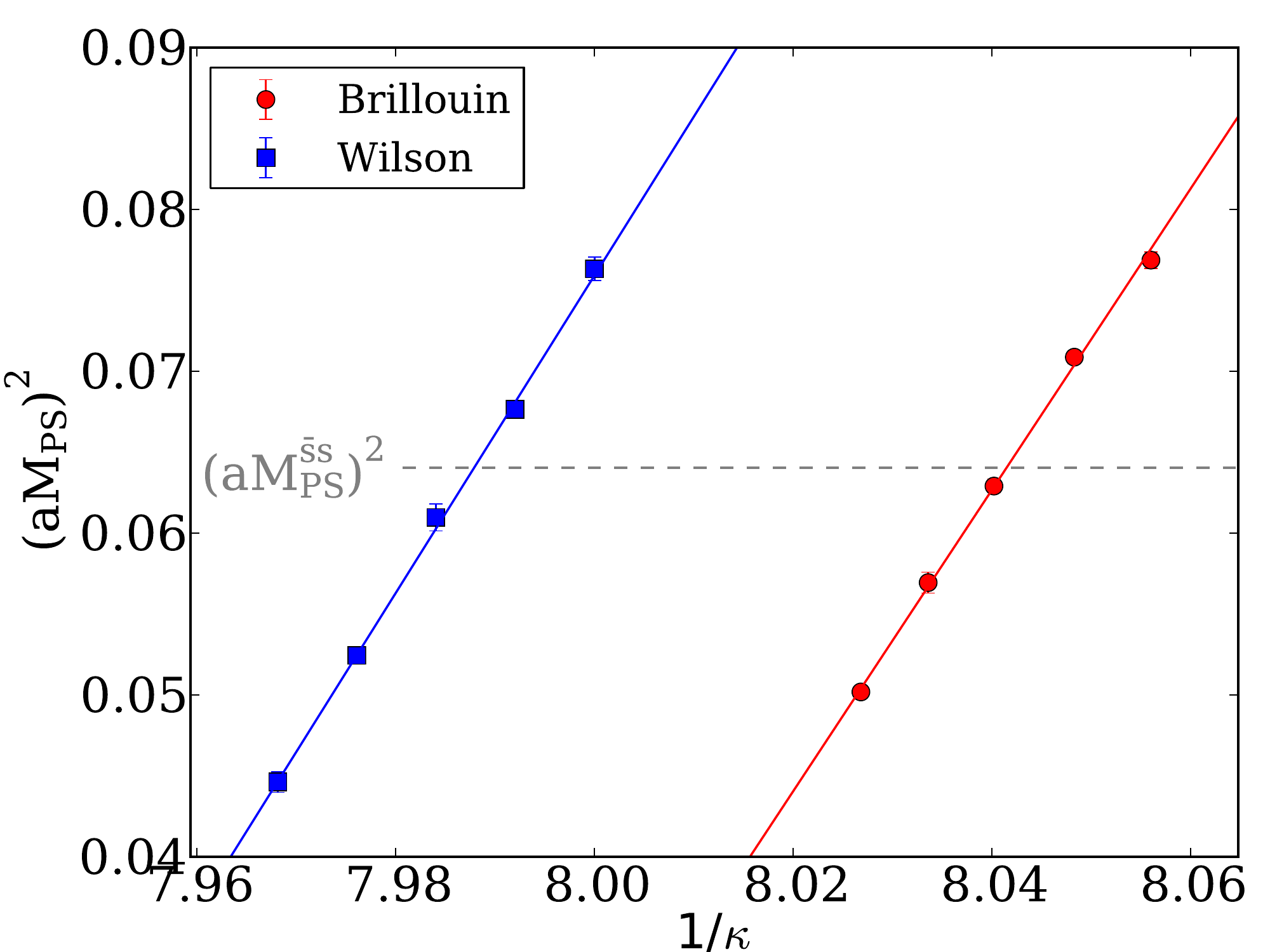}
\caption{\label{fig:tune}\sl
Left: Tuning of $\ka_l$ to achieve $(aM_P)^2=0.01069$ with either action.
Right: Similar tuning of $\ka_s$ to achieve $(aM_P)^2=0.06404$.}
\end{figure}

For the light quark we demand the pion mass to be the same (in lattice units)
as in the sea, that is $(aM_P)^2=0.01069$, given the information provided
beneath (\ref{ensemble}).
We solve the Dirac equation for a few $\ka$-values in the vicinity of the
suspected target value, and interpolate them linearly to obtain the desired
$\ka_l$, as shown in the left panel of Fig.\,\ref{fig:tune}.
The results read
\beq
\ka_l^\mr{Bri}=0.125249
\;,\qquad
\ka_l^\mr{Wil}=0.126146
\;.
\label{tuning_l}
\eeq
For the strange quark we use the scale provided below (\ref{ensemble}) and the
value $M_{s\bar{s}}=0.6858(7)\GeV$ which follows via $\sqrt{2\Mka^2-\Mpi^2}$
from the isospin-averaged and electromagnetically corrected masses
$\Mpi=134.8(3)\MeV$ and $\Mka=494.2(5)\MeV$ \cite{Colangelo:2010et}.
This gives the target value $(aM_P)^2=0.06404$, and a similar interpolation
procedure, shown in the right panel of Fig.\,\ref{fig:tune}, yields
\beq
\ka_s^\mr{Bri}=0.1251902
\;,\qquad
\ka_s^\mr{Wil}=0.1243560
\;.
\label{tuning_s}
\eeq
For the charm quark we proceed analogously to the strange case, except that
we now use the value $M_{c\bar{c}}=2.9810(11)\GeV$ from PDG
\cite{Nakamura:2010zzi}.
This yields the target value $(aM_P)^2=1.210$, and with essentially the same
kind of procedure we find
\beq
\ka_c^\mr{Bri}=0.112336
\;,\qquad
\ka_c^\mr{Wil}=0.112513
\;.
\label{tuning_c}
\eeq

\begin{figure}
\includegraphics[width=8.4cm]{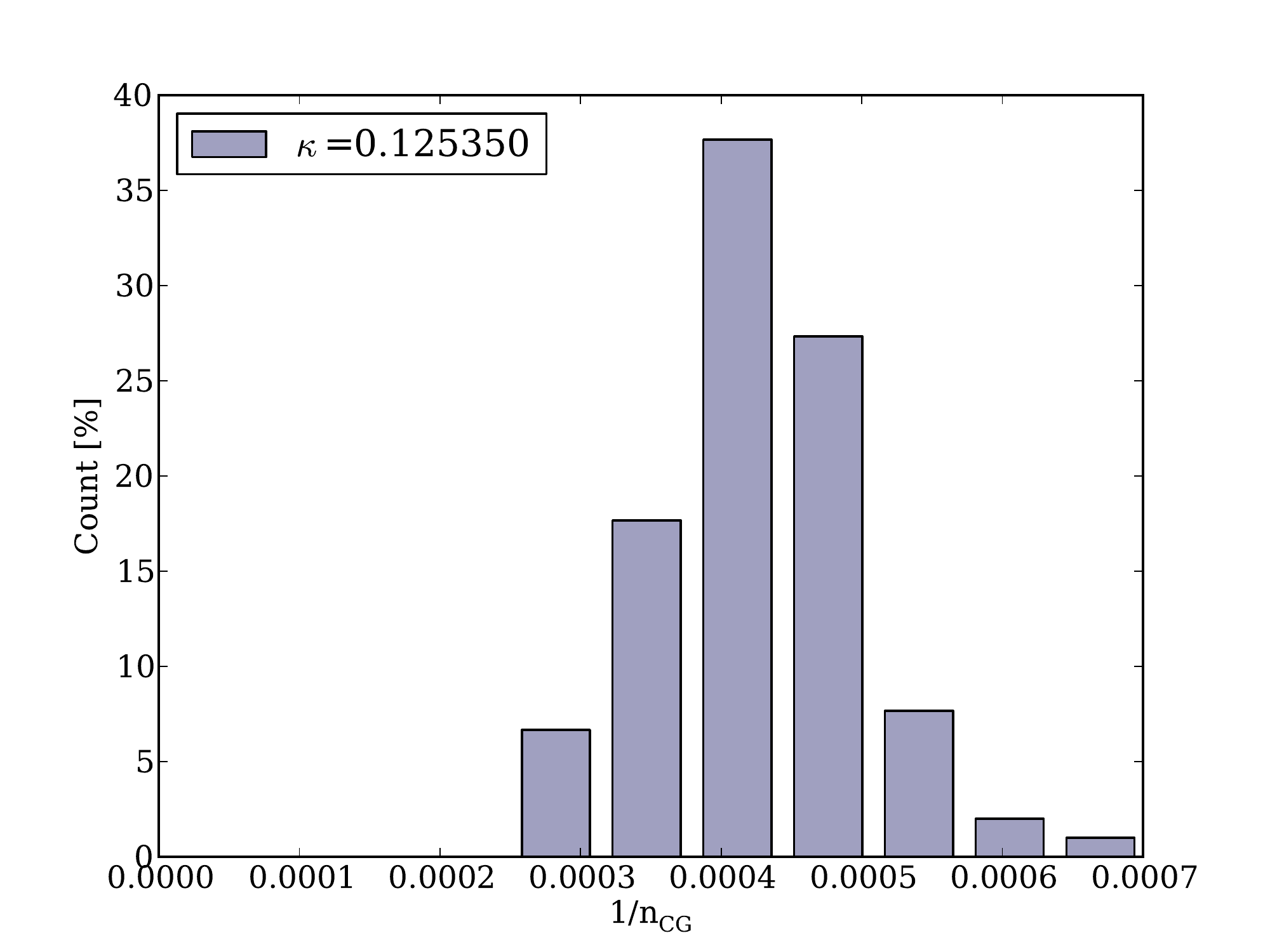}
\includegraphics[width=8.4cm]{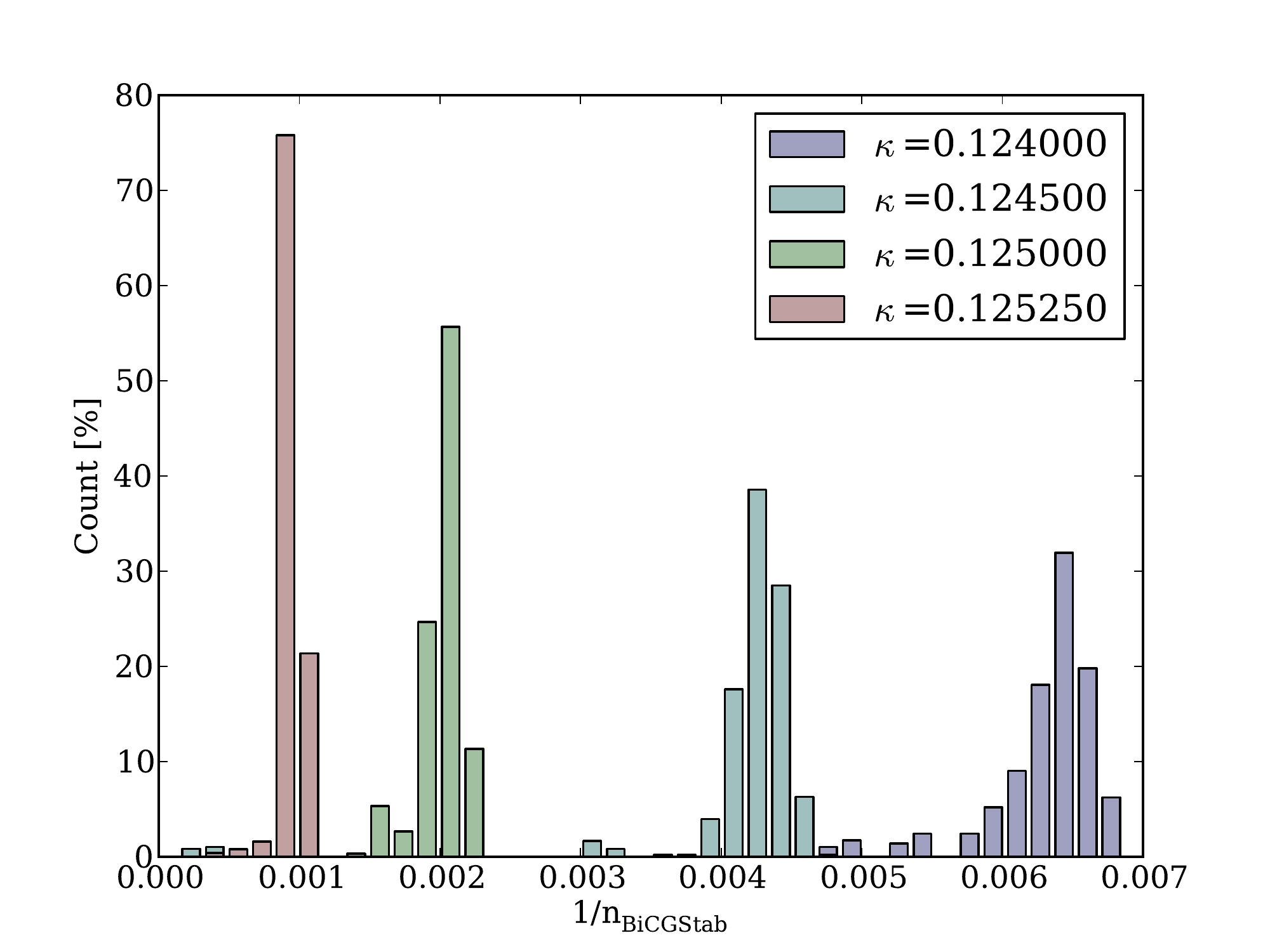}
\caption{\label{fig:itcount}\sl
Iteration count of the inversions with the trial $\ka_l$ of
Fig.\,\ref{fig:tune} for the Brillouin action, using the CG (left) and the
BiCGstab (right) algorithm for the lightest and all but the lightest masses,
respectively. In each case $O(25)$ configurations are used.}
\end{figure}

A careful look at the left panel of Fig.\,\ref{fig:tune} reveals that one of
our trial $\ka_l$-values for the Brillouin action happens to be rather light;
for this point we find $a\Mpi=0.067(5)$, tantamount to $\Mpi\simeq180\MeV$.
Following \cite{DelDebbio:2005qa,Durr:2010aw} we monitor the inverse iteration
count of the solver (which is a proxy for the smallest eigenvalue of $D\dag D$)
to make sure that we do not run into an ``exceptional configuration'' problem.
The results of this monitoring are displayed in Fig.\,\ref{fig:itcount}.
Even for the lightest quark mass the distribution is roughly Gaussian, and the
origin is 3 to 4 standard deviations away from the median.
This shows that even in this strongly non-unitary regime (where $\Mpi^\mr{val}$
is about $100\MeV$ lighter than $\Mpi^\mr{sea}$) the Brillouin operator can be
safely inverted.
We consider this an encouraging sign of the great stability of our action
against fluctuations of the small eigenmodes, and think that this stability has
the potential to render the Brillouin operator a cheap alternative to overlap
or domain-wall fermions.


\section{Meson and baryon dispersion relations \label{sec:4}}


With the tuned $\ka_l$, $\ka_s$, $\ka_c$ of (\ref{tuning_l}, \ref{tuning_s},
\ref{tuning_c}) in hand, we are now in a position to study the dispersion
relation $E^2=E^2(\mb{p}^2)$ for mesons and baryons composed of either
Brillouin or Wilson fermions.

To this end we consider two-point correlators of the form
\bea
C_M(t,\mb{p})=&{}&
\sum_{\mb{x}}\<J\!_{M}(\mb{x},t)\bar{J}\!_M(\mb{0},0)\>\,e^{\ri\mb{px}}
\\
C_B^\pm(t,\mb{p})=&\Trc\,\frac{1}{2}(1\pm\ga_4)&
\sum_{\mb{x}}\<J_B(\mb{x},t)\,\bar{J}_B(\mb{0},0)\>\,e^{\ri\mb{px}}
\label{corr_baryon}
\eea
where $M$ and $B$ define the quantum numbers of the meson or baryon,
respectively.
For non-zero momentum the correct parity projector in (\ref{corr_baryon}) would
read $\frac{1}{2}(1\pm\frac{E}{M}\ga_4)$ \cite{Lee:1998cx,Leinweber:2004it},
but we stay with the simpler form, since we are always interested in the
lower-mass parity partner (which requires no projection at all).
Finally, the sink and the source in (\ref{corr_baryon}) contain an uncontracted
spinor index, say $\mu$ and $\nu$.
The projection to a definite spin can be done with \cite{Bowler:1996ws}
\bea
(P^{3/2})_{\mu\nu}&=&\de_{\mu\nu}
-\frac{1}{3}\gam\gan-\frac{1}{3p^2}(p\!\!\!/\gam p_\nu-p_\mu\gan p\!\!\!/)
\\
(P^{1/2}_{11})_{\mu\nu}&=&\frac{1}{3}\gam\gan
-\frac{1}{p^2}p_\mu p_\nu+\frac{1}{3p^2}(p\!\!\!/\gam p_\nu-p_\mu\gan p\!\!\!/)
\eea
and similar expressions for $P_{12,21,22}^{1/2}$ as given in eqn.\,(9) of
\cite{Bowler:1996ws}.
For $\mb{p}=0$ they simplify to
\bea
(P^{3/2})_{ij}&=&\de_{ij}-\frac{1}{3}\ga_i \ga_j
\\
(P^{1/2}_{11})_{ij}&=&
\frac{1}{3}\ga_i \ga_j
\eea
which are easy to implement and good enough for many purposes (we are always
interested in the lowest-mass state).
For details of (more advanced) spin projection see \cite{Edwards:2011jj}.

\begin{figure}[!tb]
\includegraphics[width=8.4cm]{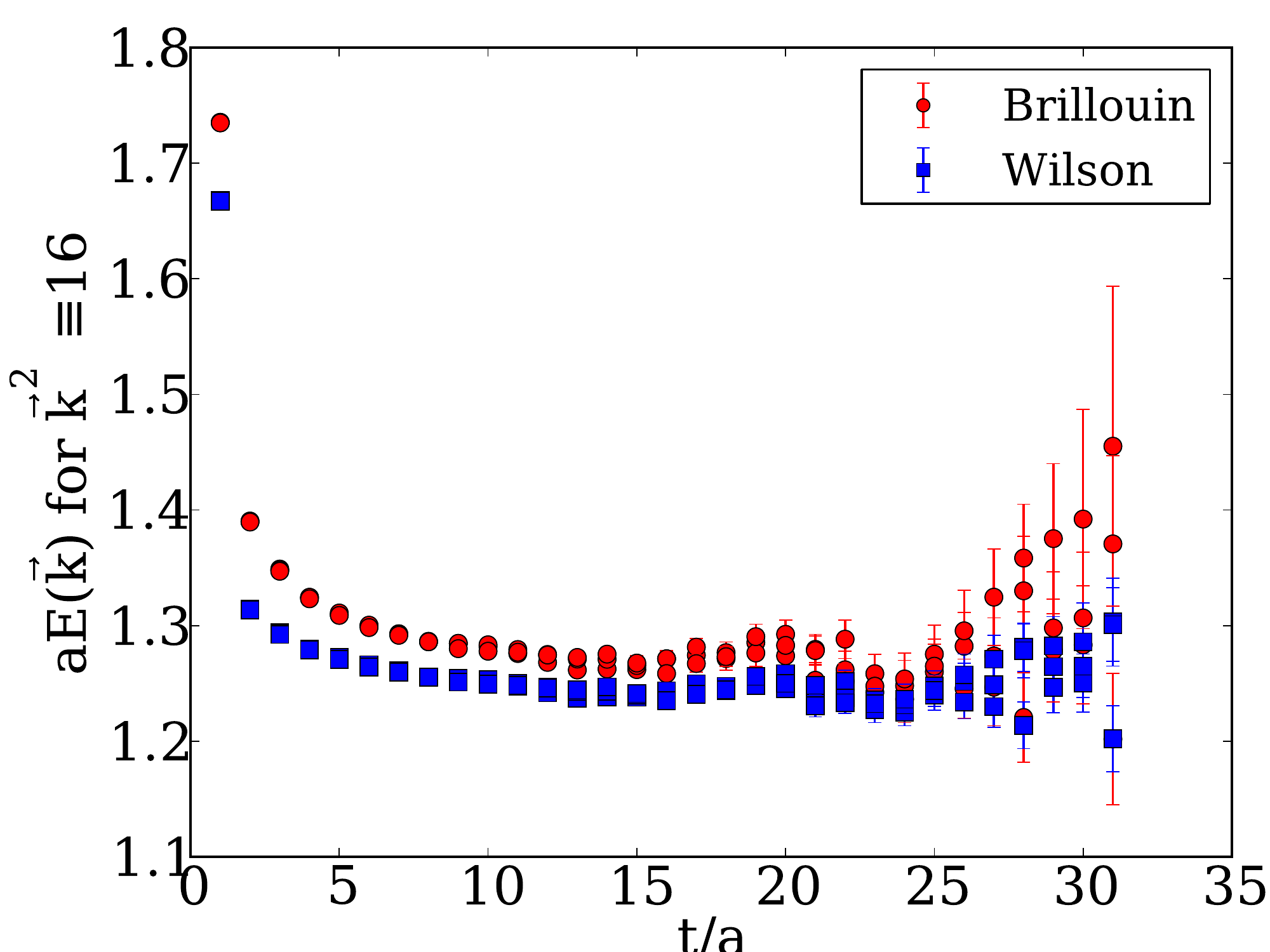}
\includegraphics[width=8.4cm]{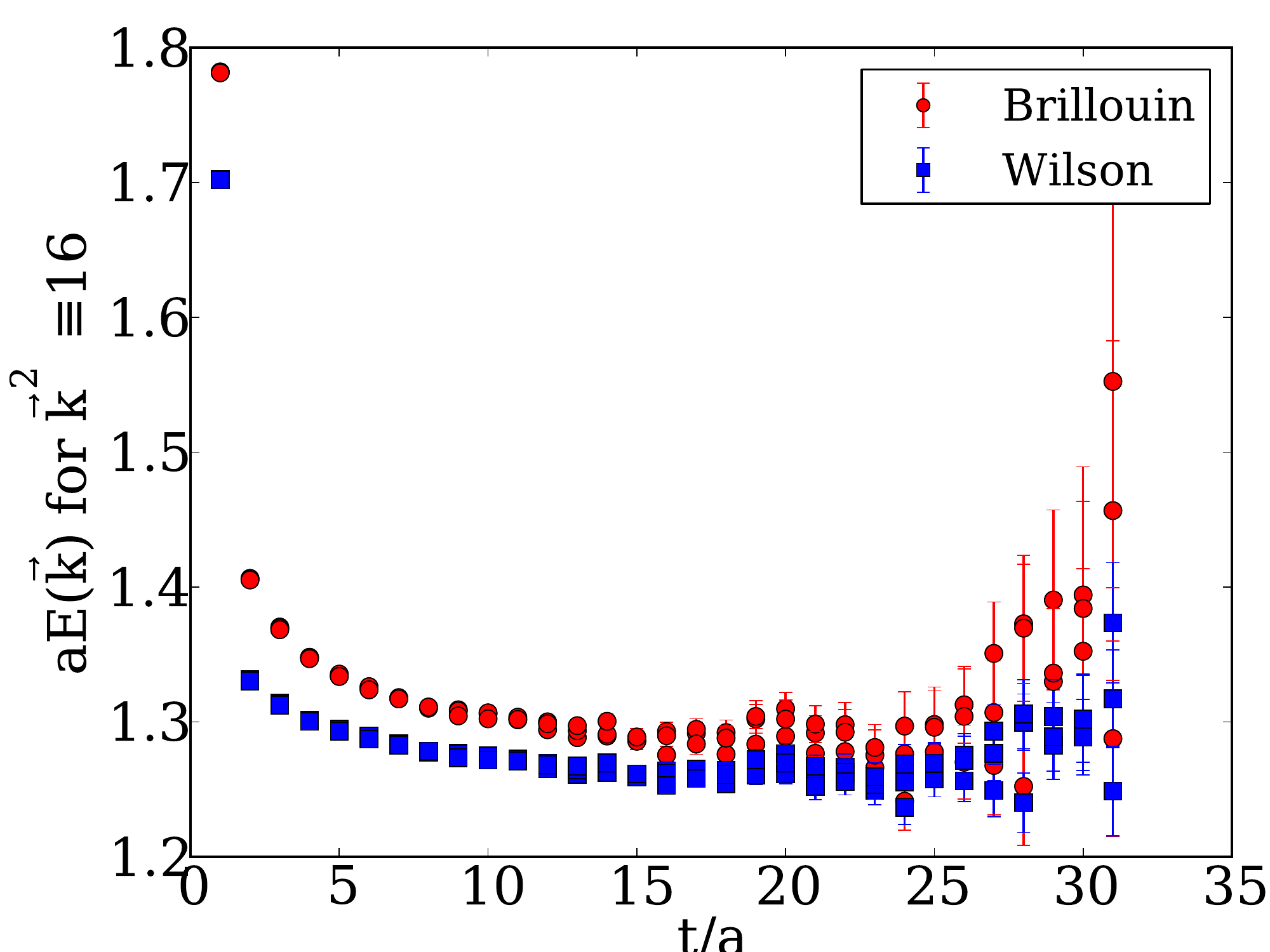}
\\
\includegraphics[width=8.4cm]{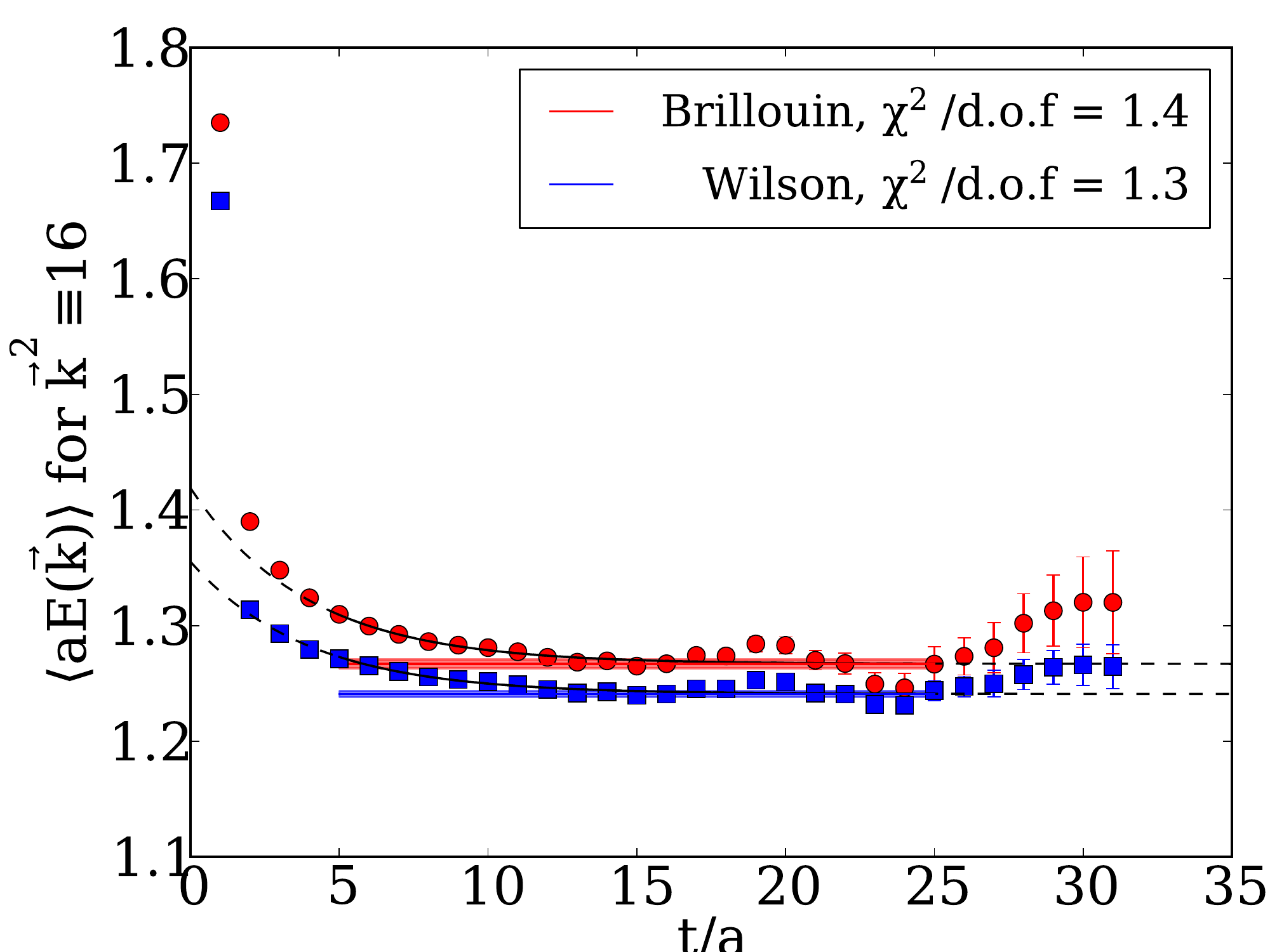}
\includegraphics[width=8.4cm]{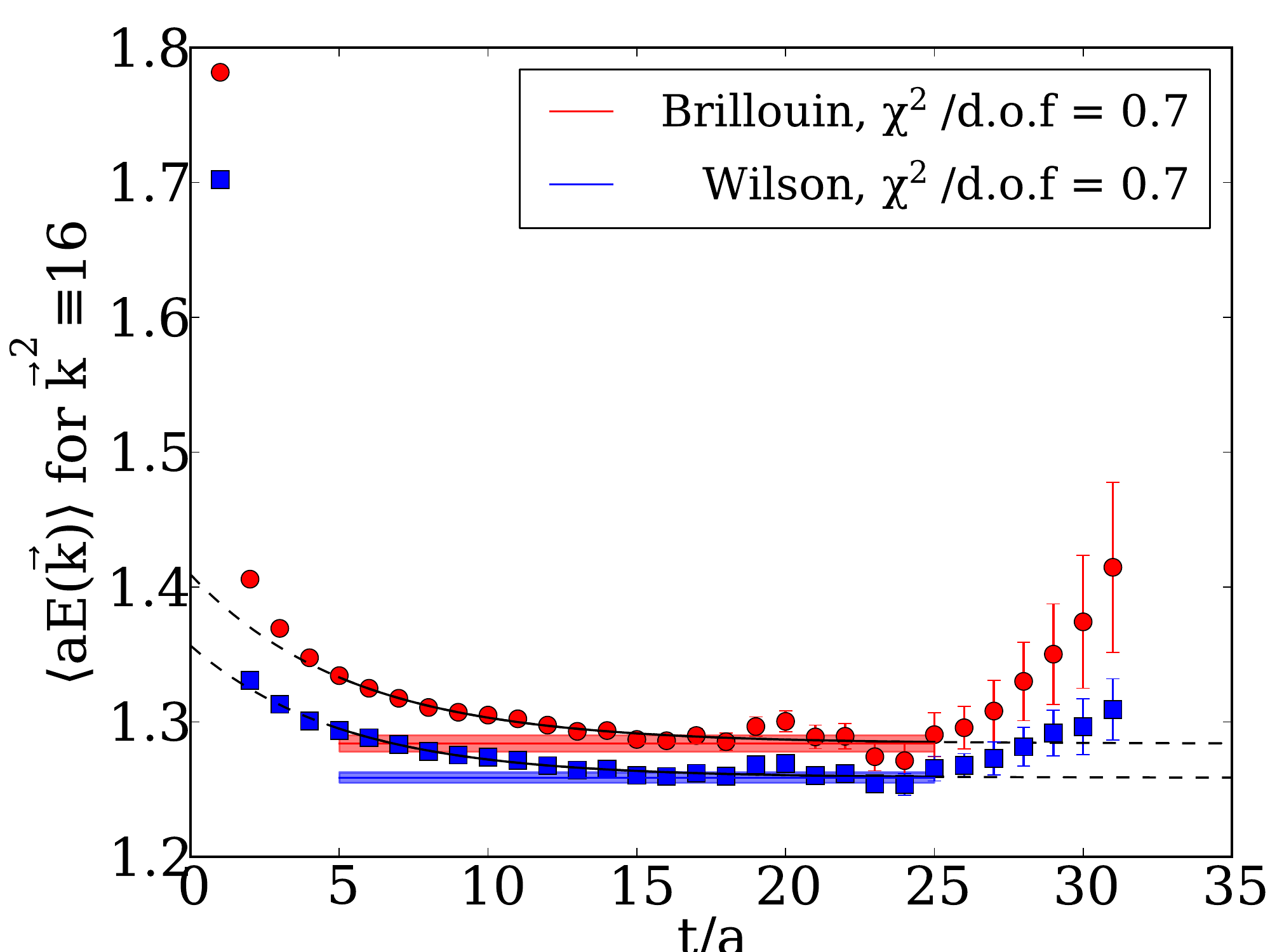}
\caption{\label{fig:average}\sl
Effective mass plots of the pseudoscalar (left) and vector (right) $c\bar{c}$
meson before (top) and after (bottom) averaging over the
$\mb{k}$-configurations that contribute to $\mb{k}^2=16$.}
\end{figure}

Having defined the correlation functions, we can now move on to the dispersion
relations.
We consider spatial momenta $\mb{p}=2\pi/L\cdot\mb{k}$ with
$0\leq\mb{k}^2\leq20$.
Thanks to the cubic symmetry among the spatial directions, in general several
$\mb{k}$-configurations contribute to a given $\mb{k}^2$.
An effective mass plot before and after taking an average over the various
contributions is shown in Fig.\,\ref{fig:average}, for the pseudoscalar
and vector meson, in the case of $\mb{k}^2=16$.
We see no big discrepancies before the average is taken, hence the averaging
seems justified.

\begin{figure}
\centering
\includegraphics[width=15cm]{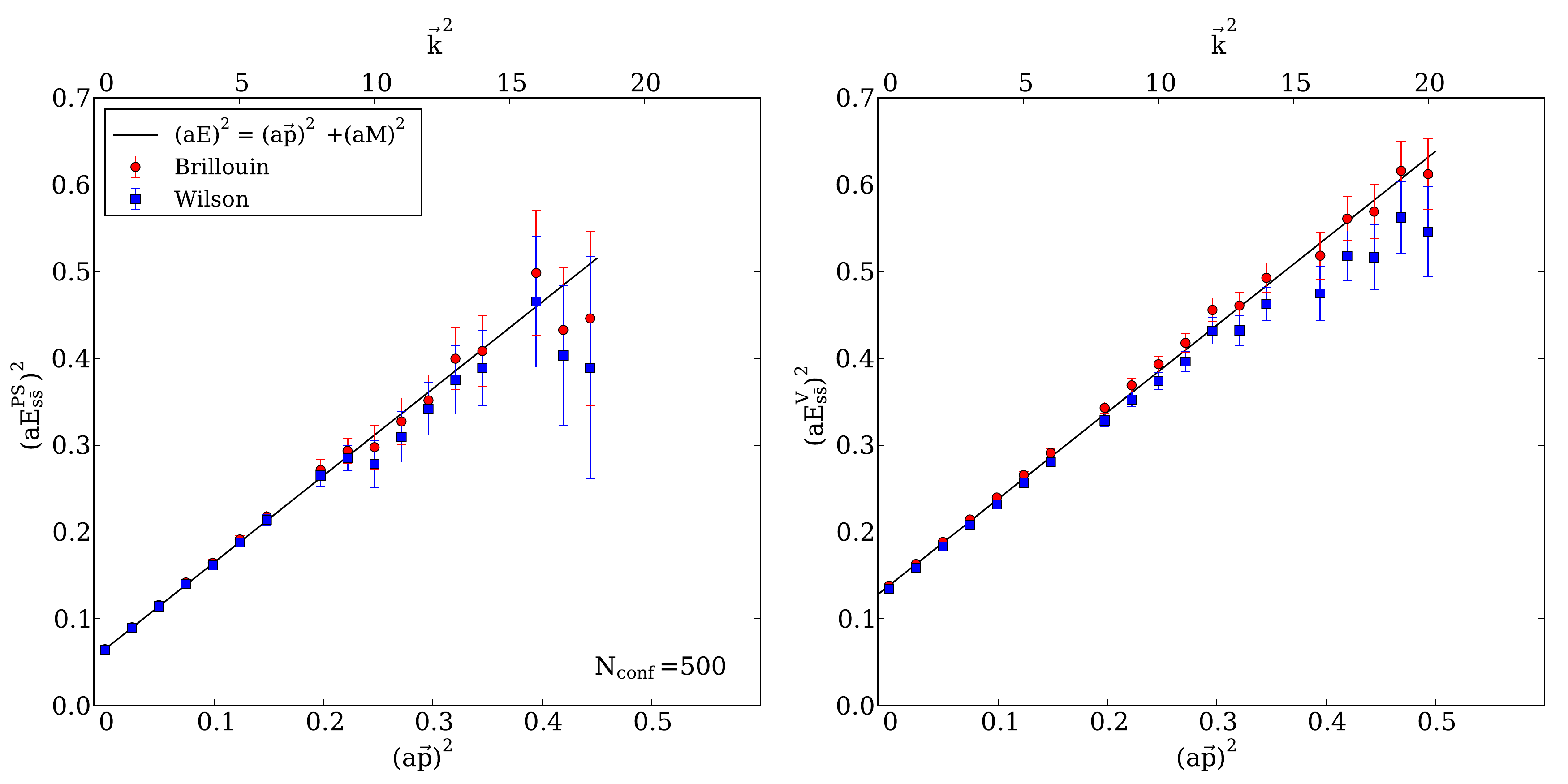}\\
\includegraphics[width=15cm]{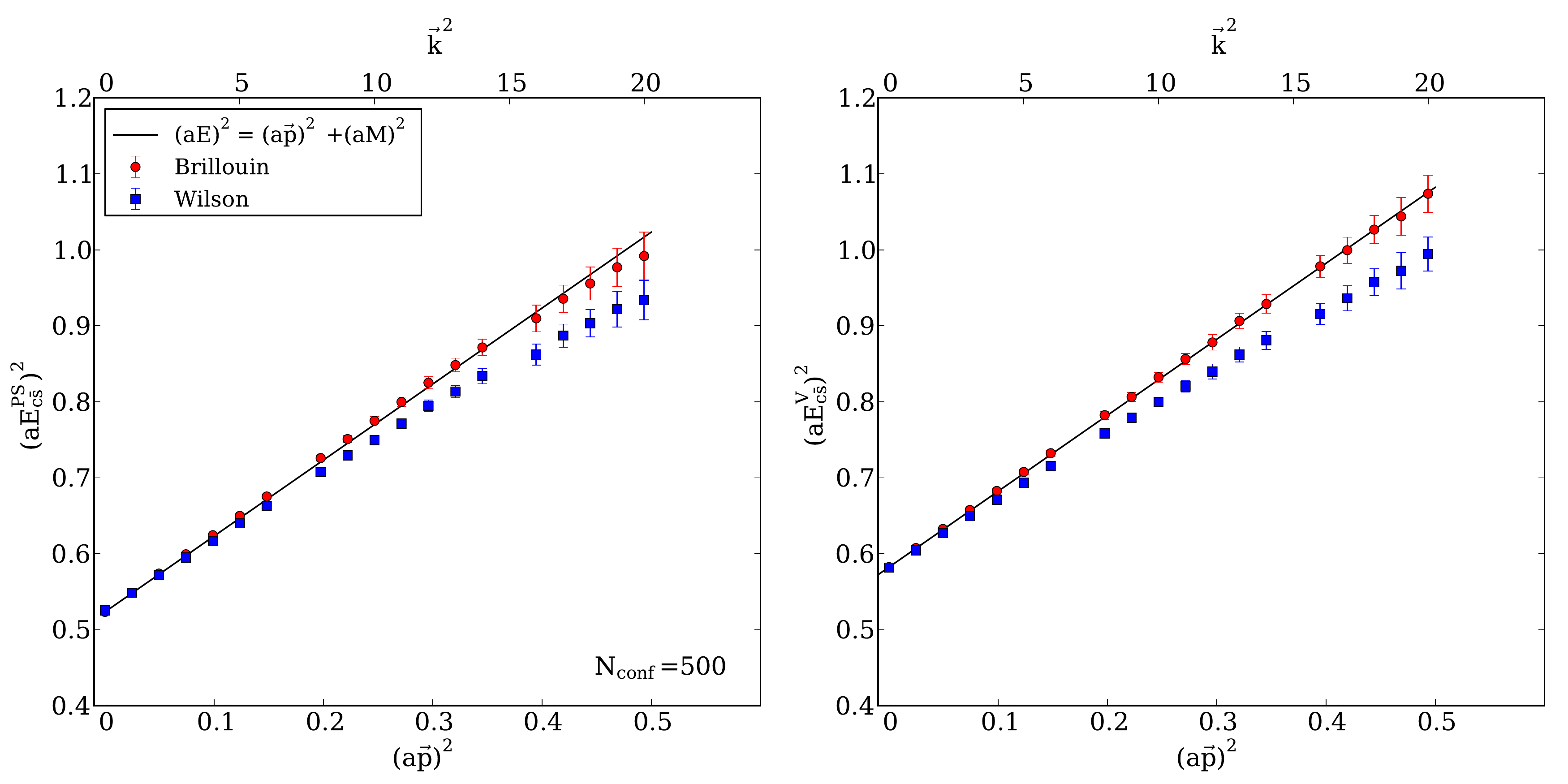}\\
\includegraphics[width=15cm]{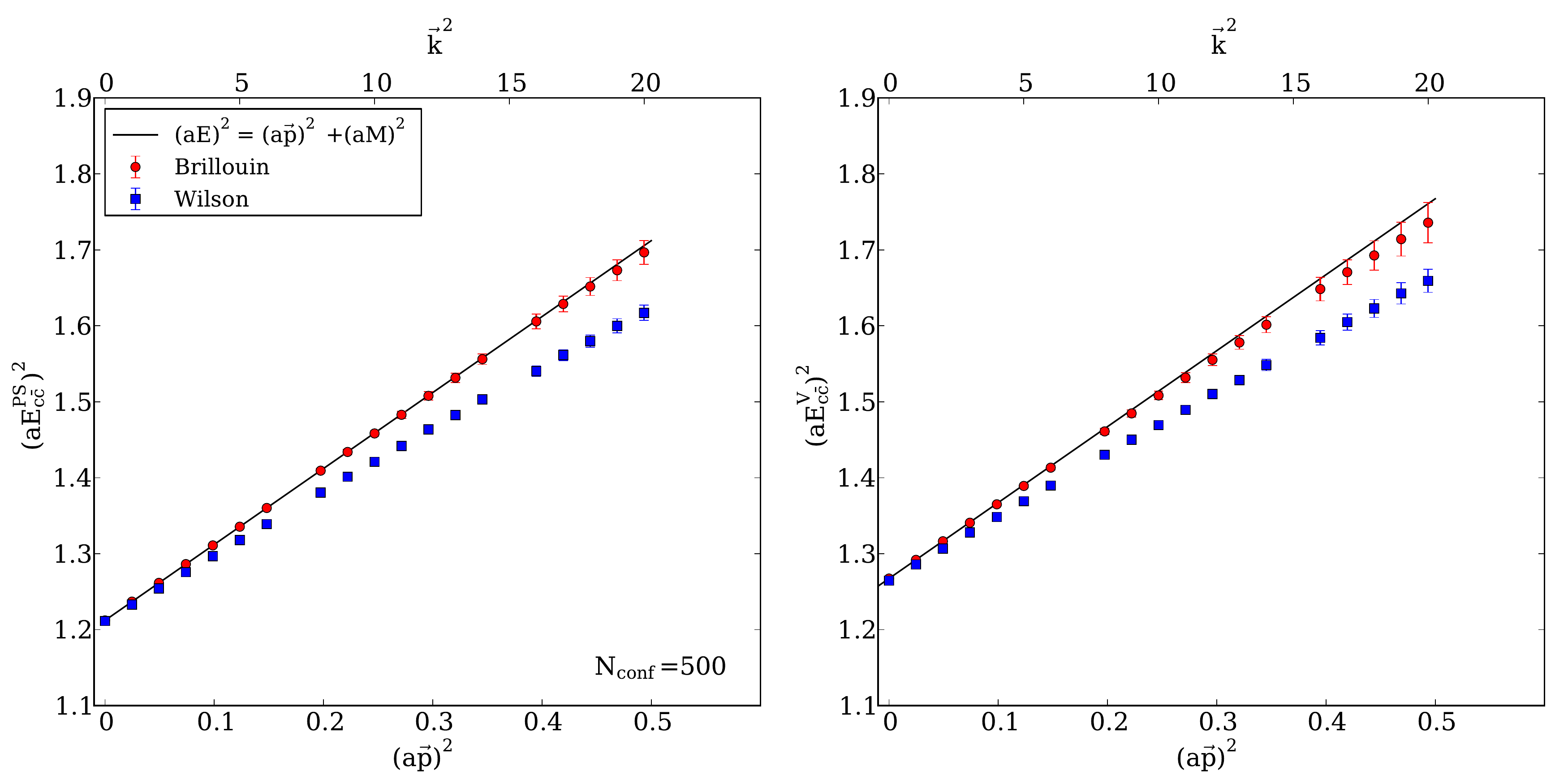}\vspace{-2pt}
\caption{\label{fig:disp_mesons}\sl
Dispersion relations for the pseudoscalar (left) and vector (right) meson with
$s\bar{s}$ (top), $s\bar{c}$ (middle), $c\bar{c}$ (bottom) quark content. The
black line shows the relativistic $E^2=\mb{p}^2+M^2$.}
\end{figure}

Repeating this procedure for all $\mb{k}^2$ yields the data presented in
Fig.\,\ref{fig:disp_mesons}.
We show the dispersion relations for the pseudoscalar and the vector meson
with quark content $s\bar{s}$ (top), $s\bar{c}$ (middle), $c\bar{c}$ (bottom).
Evidently, for the heavier masses there is a significant difference between the
Wilson data (red circles) and those with the Brillouin discretization (blue
squares).
The full black line is not a fit, but the relativistic dispersion relation
$E^2=\mb{p}^2+M^2$, starting from the first data-point (where the two actions
were tuned to yield the same result for the $s\bar{s}$ and $c\bar{c}$
pseudoscalar states, but not for the remaining four states).
This line shows that the discretization effects are induced by the Wilson
action; within statistical errors the Brillouin data are free from such
effects.

\begin{figure}[!tb]
\centering
\includegraphics[width=15cm]{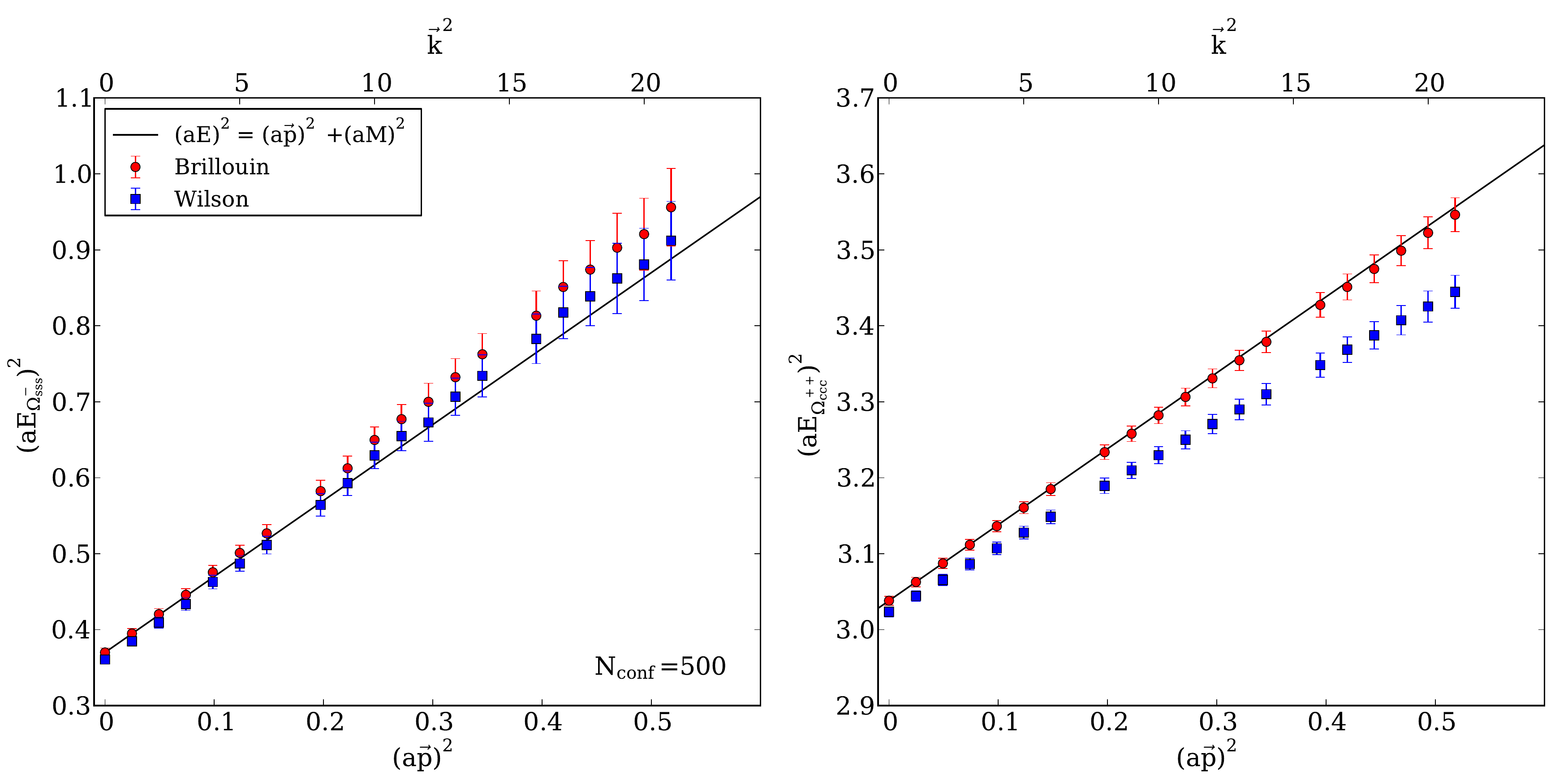}\\
\includegraphics[width=15cm]{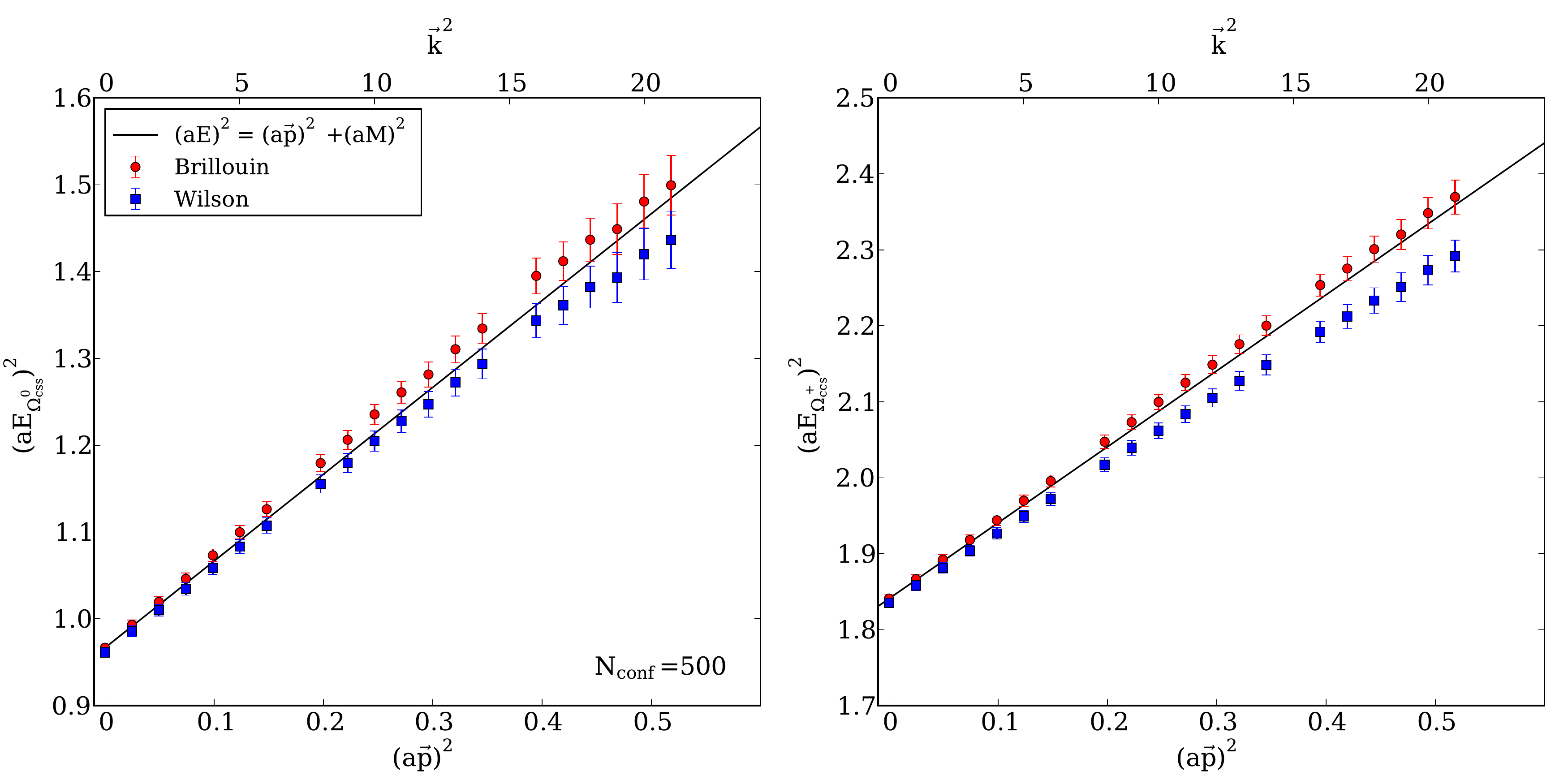}
\caption{\label{fig:disp_omegas}\sl
Dispersion relations for the decuplet-type $\Omega_{sss}^-$,
$\Omega_{ccc}^{++}$ (top left and right) and the octet-type $\Omega_{css}^0$,
$\Omega_{ccs}^+$ (bottom left and right). The black line shows the relativistic
$E^2=\mb{p}^2+M^2$.}
\end{figure}

Similarly, we can work out the dispersion relations for baryons.
The data for the ``decuplet-type'' states $\Omega_{sss}^-$,
$\Omega_{ccc}^{++}$ [which form a 20-plet under SU(4)], as well as for the
``octet-type'' states $\Omega_{css}^0$, $\Omega_{ccs}^+$ [which form a
$20'$-plet under SU(4)] are shown in Fig.\,\ref{fig:disp_omegas}.
To make it clear which states we consider, the interpolating fields of both the
20-plet and the $20'$-plet are listed in the appendix.
Again, the full black line is not a fit but the relativistic dispersion
relation $E^2=\mb{p}^2+M^2$, and we stress that no further tuning of
$\ka$-values was performed.
Just like in the meson case, we see no significant difference in the regime of
the physical strange quark mass (which explains why we refrain from looking at
even lighter $\ka$ values, as such data would just be more noisy).
However, with every strange quark that is replaced by a charm quark, the
difference becomes more pronounced, up to the point where the dispersion
relation of the $\Omega_{ccc}^{++}$ is seriously distorted with Wilson
fermions, but relativistically correct with Brillouin fermions.


\section{Masses of multiply charmed baryons \label{sec:5}}


As a byproduct of our investigation, and with the goal of spurring further
improvements, we can quote the masses of the $\Omega$ baryons considered in
the previous section.
Let us emphasize that this activity is based on the single ensemble
(\ref{ensemble}).
We will try to minimize the systematic effects on the numbers given below, but
there is no way of reliably assessing the size of such systematic effects,
given the data that we have.

\begin{figure}[!tb]
\centering
\includegraphics[width=15cm]{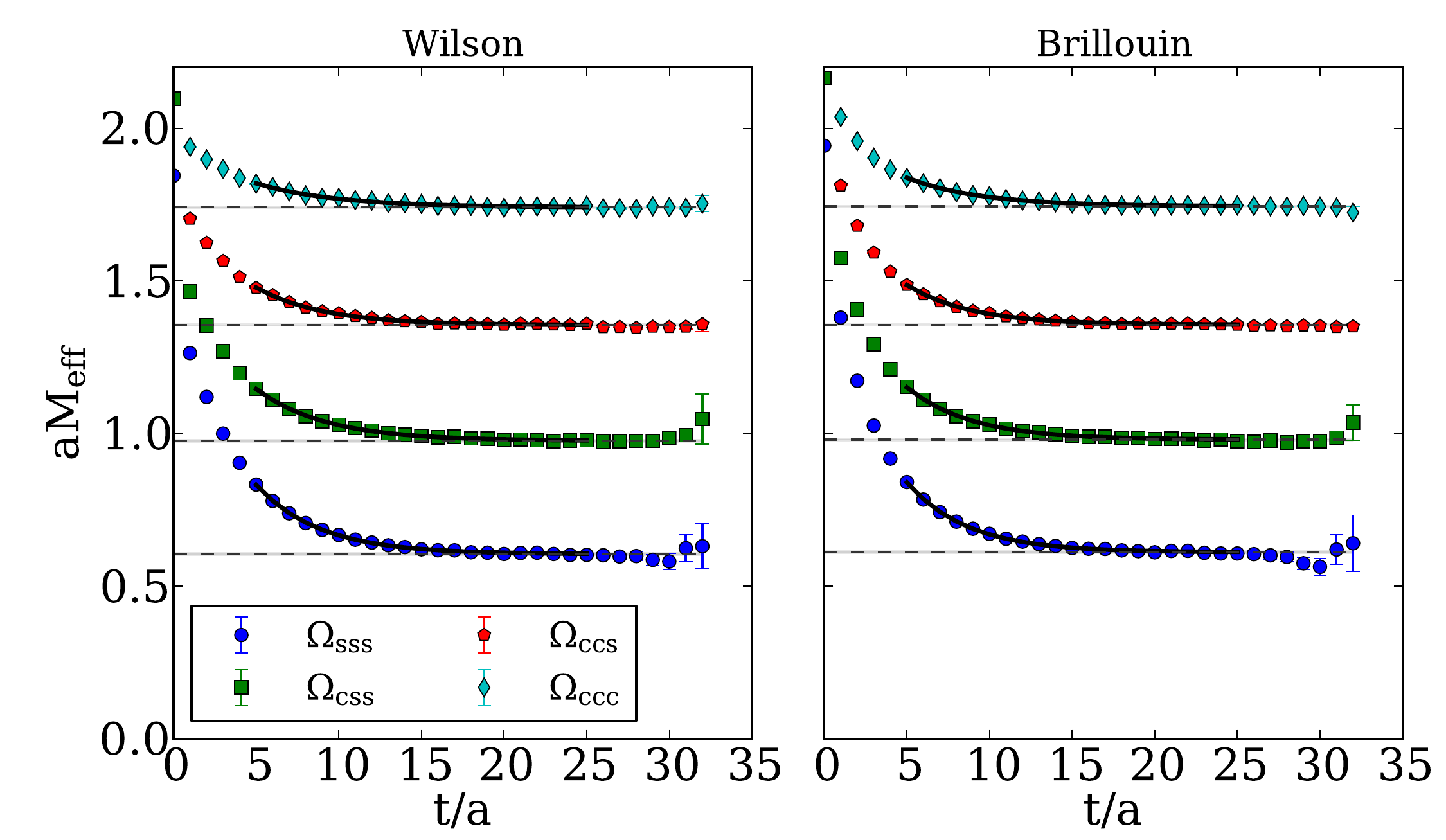}
\caption{\label{fig:omega_meff_direct}\sl
Effective masses of the $\Omega_{sss}$, $\Omega_{ccc}$ (decuplet-type states)
and $\Omega_{ssc}$, $\Omega_{scc}$ (octet-type states) at $\mb{p}=0$ for the
Wilson (left) and Brillouin (right) action.}
\end{figure}

\begin{figure}[!tb]
\centering
\includegraphics[width=15cm]{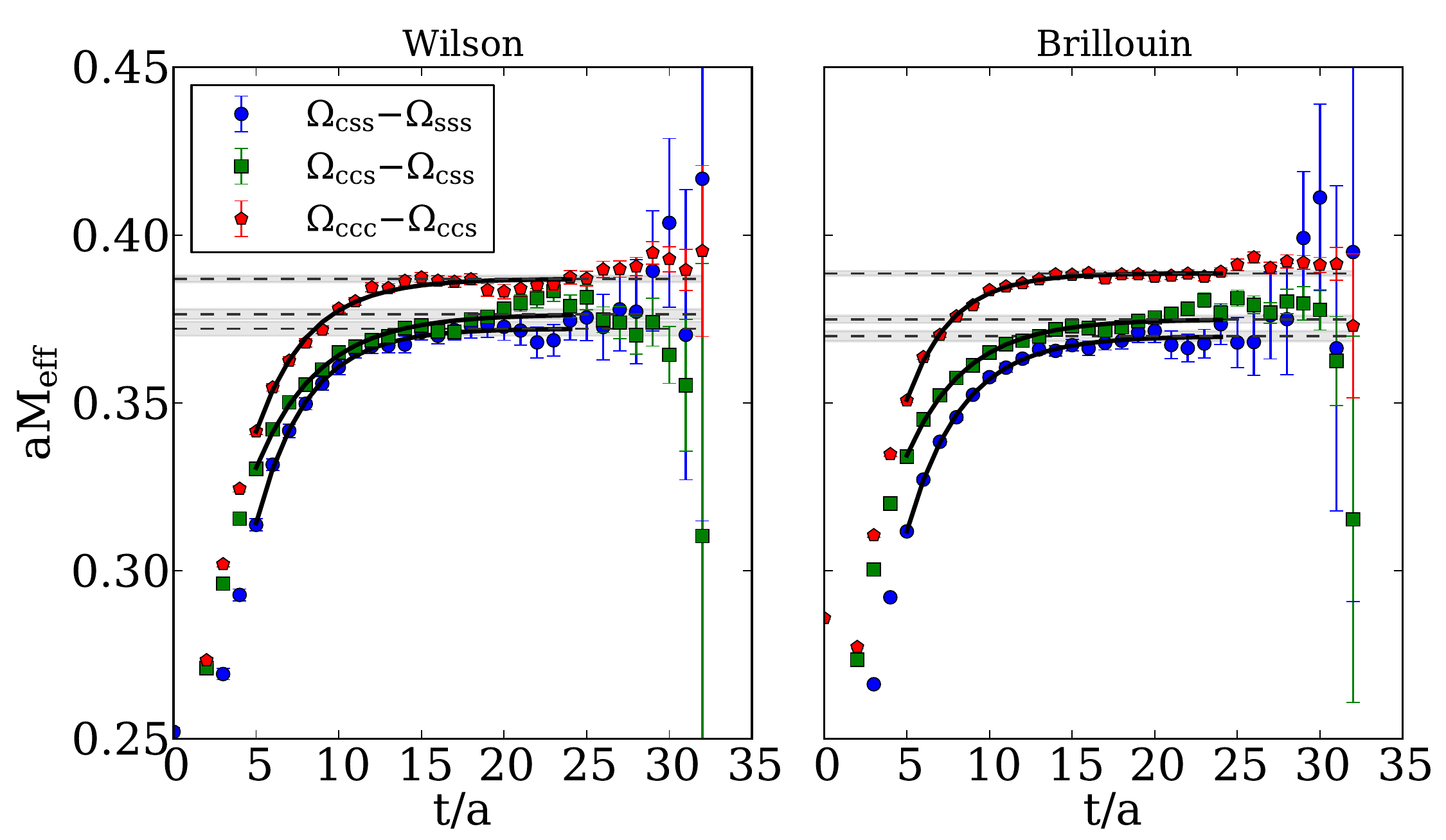}
\caption{\label{fig:omega_meff_ratios}\sl
Same as Fig.\,\ref{fig:omega_meff_direct}, but for the splittings
$M_{\Omega_{css}}\!-\!M_{\Omega_{sss}}$ and
$M_{\Omega_{ccs}}\!-\!M_{\Omega_{css}}$ and
$M_{\Omega_{ccc}}\!-\!M_{\Omega_{ccs}}$.}
\end{figure}

\begin{table}[!tb]
\centering
\begin{tabular}{|ccc|cc|}
\hline
combination & Wilson & Brillouin \\
\hline
$(\Omega_{css}-\Omega_{sss})/\Omega_{sss}$ & 0.615(6)& 0.605(5) \\
$(\Omega_{ccs}-\Omega_{css})/\Omega_{sss}$ & 0.622(5)& 0.613(4) \\
$(\Omega_{ccc}-\Omega_{ccs})/\Omega_{sss}$ & 0.639(4)& 0.636(4) \\
\hline
$(\Omega_{ccs}-\Omega_{css})/\Omega_{css}$ & 0.386(2)& 0.383(2) \\
$(\Omega_{ccc}-\Omega_{ccs})/\Omega_{ccs}$ & 0.286(1)& 0.287(1) \\
\hline
\end{tabular}
\caption{\label{tab:omegasplittings}\sl
Relative mass splittings between the various $\Omega$ states, as determined
with the Wilson (left) and the Brillouin (right) action. Here, the symbol
$\Omega_{xyz}$ is meant as a shorthand for $M_{\Omega_{xyz}}$.}
\end{table}

Let us begin by showing the effective mass plots of the four states
$\Omega_{sss}^-$, $\Omega_{css}^0$, $\Omega_{ccs}^+$, $\Omega_{ccc}^{++}$ in
Fig.\,\ref{fig:omega_meff_direct}.
This is at $\mb{p}=0$, and we see no significant difference between the Wilson
and the Brillouin action.
Let us recall that the tuning was done in the meson sector, so this is a first
non-trivial observation.
To minimize the systematics it is usually a good idea to look at mass
splittings, and to form dimensionless ratios.
The former trick mitigates the effect of excited states contaminations and
possible finite volume effects, the latter one reduces the sensitivity to the
overall scale.
The effective mass plots for all adjacent mass splittings, based on ratios of
correlators like $\<C_{css}\>/\<C_{sss}\>$, are shown in
Fig.\,\ref{fig:omega_meff_ratios}.
Even at this level of zoom, the quality of the data appears rather good, and we
see only very mild differences between the Wilson and the Brillouin data.
We list the relative mass splittings (normalized both with the lower one of the
two states involved and with the $\Omega^-=\Omega_{sss}$ base state) in
Tab.\,\ref{tab:omegasplittings}.



Regarding phenomenological numbers, we should first mention that the
experimental mass of the $\Omega^-$ (i.e.\ the $J^P=3/2^+$, $c=0$, $s=3$ state
in Tab.\,\ref{tab:dec}) is $1672.45(29)\MeV$, and the mass of the $\Omega_c^0$
(i.e.\ the $J^P=1/2^+$, $c=1$, $s=2$ state in Tab.\,\ref{tab:oct}) is
$2697.5(2.6)\MeV$ \cite{Nakamura:2010zzi}.
Hence, with the ratios listed in Tab.\,\ref{tab:omegasplittings}, we can only
compare the mass of the $\Omega_c^0$ baryon to experiment, but not the masses
of the $\Omega_{cc}^+$ (i.e.\ the $J^P=1/2^+$, $c=2$, $s=1$ state in
Tab.\,\ref{tab:oct}) and of the $\Omega_{ccc}^{++}$ (i.e.\ the $J^P=3/2^+$,
$c=3$, $s=0$ state in Tab.\,\ref{tab:dec}).
We shall use the first three lines in Tab.\,\ref{tab:omegasplittings}, taking
the Brillouin number as our central value and the average difference to the
Wilson number (0.007) as a uniform estimate of the systematic error.
Adding all errors in quadrature yields
\bea
 M_{\Omega_{c}^{0}}&=&2685(15)\MeV
\label{postdiction}
\\
 M_{\Omega_{cc}^{+}}&=&3711(20)\MeV
\label{prediction_one}
\\
 M_{\Omega_{ccc}^{++}}&=&4774(24)\MeV
\label{prediction_two}
\eea
and we emphasize that these errors do not include the effect of the (missing)
limits $a\to0$ and $\Mpi^\mr{sea}\to134.8\MeV$.
Nevertheless, (\ref{postdiction}) is consistent with experiment, albeit
with a large error.
From these numbers it appears that the actual splitting is very close to
equidistant, a notion which is also conveyed by
Fig.\,\ref{fig:omega_meff_ratios}.

We refrain from comparing our numbers to similar results in the recent
literature on charm physics on the lattice \cite{Bowler:1996ws,Lewis:2001iz,
Mathur:2002ce,Flynn:2003vz,Na:2008hz,Liu:2009jc,Lin:2010wb,Mohler:2011ke,
Namekawa:2011wt,Lin:2011ti,Alexandrou:2012xk,Briceno:2012wt}.
We rather like to add that what is really called for, in our opinion, is a
complete study with a reasonable assessment of all systematics involved, that
is with the continuum limit taken, with an interpolation or extrapolation to
the physical values of $\Mpi$ and $\Mka$ in the sea, and with an extrapolation
to infinite box volume.


\section{Summary \label{sec:6}}


The goal of this work was to test whether a significant difference in meson
and baryon dispersion relations is seen, depending on whether such a composite
state is built from Wilson or Brillouin fermions.
The main result is that for standard lattice spacings ($a^{-1}=2-3\GeV$) this
is not the case if all quarks are at most as heavy as the physical strange
quark, but significant differences become visible if one or several quarks are
in the range of the physical charm quark mass -- see Fig.\,\ref{fig:disp_mesons}
for mesons and Fig.\,\ref{fig:disp_omegas} for baryons.

We should add that --~even for heavy quarks~-- there is an alternative in case
one is willing to give up on Lorentz invariance \cite{ElKhadra:1996mp}.
In the original version of this Fermilab method new anisotropy parameters were
introduced which may be tuned to get the slope (``speed of light'') of the
pseudoscalar dispersion relation correct.
Nowadays it is more common to focus on the non-relativistic behavior
$E(\mb{p})=M_1+\mb{p}^2/(2M_2)+...$ and to choose $\ka$ such that the
kinetic mass $M_2$ is correct.
In the first version the price to pay is the added expense, in terms of CPU and
human time, to tune the parameters to sufficient precision (which is usually
not too hard in a quenched context, but the issue will be aggravated by
renormalization effects, once the charm quark is unquenched).
In the second version only mass splittings among states with the same number of
charm quarks may be considered, such that the rest mass $M_1$ drops out.
By contrast the relativistic setup of the Brillouin action requires no tuning
and no compromises to be made on the set of calculable observables; in our view
it wins in terms of ease of use.

The Brillouin operator as proposed in \cite{Durr:2010ch} can be seen as a
low-cost approximation to the concept of ``perfect fermions''
\cite{Hasenfratz:1993sp,Bietenholz:1995cy,Hasenfratz:2000xz,Hasenfratz:2002rp}.
A recent development in the field of staggered fermions is to add a mixture of
taste-S,V,T,A,P mass terms \cite{Golterman:1984cy,Adams:2010gx,
Hoelbling:2010jw,Creutz:2010bm,deForcrand:2012bm} such that the resulting
action would have only one or two species in the continuum, and an eigenvalue
spectrum similar to the near-Ginsparg-Wilson spectrum of the Brillouin action
(see e.g.\ Fig.\,22 of Ref.\cite{Durr:2010ch}).

In summary, we reach the conclusion that the added expense, in terms of CPU
time, that the Brillouin action entails over the Wilson action is hardly
justified if one is only interested in light quark spectroscopy (this may be
different for structure functions).
On the other hand, as soon as charm quarks are involved, the Brillouin action
leads to a massive reduction of cut-off effects already in purely spectroscopic
quantities.
In particular our Fig.\,\ref{fig:disp_omegas} shows that the standard lore that
$aM$ should not exceed $1$ need not be true with Brillouin fermions; in this
figure we see no cut-off effects in the range $(aE)^2=3.0-3.5$.
All together, it seems our choice to use the Brillouin action to determine
the quark mass ratio $m_c/m_s$ in Ref.\cite{Durr:2011ed} was justified, and we
hope that this augurs well for the accuracy of our results
(\ref{prediction_one}, \ref{prediction_two}).

\bigskip\bigskip

\noindent{\bf Acknowledgments}:
We thank Constantia Alexandrou and Zoltan Fodor for useful discussion.
This work was supported in part by DFG through SFB TRR-55.
The computing resources for this project were provided by Forschungszentrum
J\"ulich GmbH through a VSR grant.

\clearpage


\section*{Appendix: Charmed baryon interpolating fields}


To avoid any confusion to which states our numbers
(\ref{postdiction}-\ref{prediction_two}) would refer to, we give a list of the
simplest baryon interpolators with charm and/or strangeness.
We follow the naming convention of PDG \cite{Nakamura:2010zzi} and group the
operators according to their transformation properties under SU(4) in flavor
space.
Throughout, $C$ denotes the charge conjugation matrix, the transposition sign
refers to spinor, and color indices are implicit, as described in the table
captions.

Overall, the states separate into a $20'$-plet of spin $1/2$ states, a
$20$-plet of spin $3/2$ states, and a $\bar{4}$-plet under SU(4).
These states are listed in Tab.\,\ref{tab:oct}, Tab.\,\ref{tab:dec}, and
Tab.\,\ref{tab:new}, respectively.






\begin{table}[!tb]
\centering
\begin{tabular}{|cc|cccc|}
\hline
charm & strange & baryon & interpolating field & $I$ & $I_z$ \\
\hline
\hline
$c=0$ & $s=0$ & $p$               & $\ep(u^T C\gaf d)u$ & $1/2$ & $+1/2$ \\
      &       & $n$               & $\ep(d^T C\gaf u)d$ & $1/2$ & $-1/2$ \\
\hline
      & $s=1$ & $\Sigma^{+}$      & $\ep(u^T C\gaf s)u$ & $1$ & $+1$ \\
      &       & $\Sigma^{0}$      & $\frac{1}{\sqrt{2}}\ep\{(u^T C\gaf s)d+(d^T C\gaf s)u\}$ & $1$ & $0$ \\
      &       & $\Sigma^{-}$      & $\ep(d^T C\gaf s)d$ & $1$ & $-1$ \\
\hline
      & $s=2$ & $\Xi^{0}$         & $\ep(s^T C\gaf u)s$ & $1/2$ & $+1/2$ \\
      &       & $\Xi^{-}$         & $\ep(s^T C\gaf d)s$ & $1/2$ & $-1/2$ \\
\hline
      & $s=1$ & $\Lambda^0$       & $\frac{1}{\sqrt{6}}\ep\{2(u^T C\gaf d)s+(u^T C\gaf s)d-(d^T C\gaf s)u\}$ & $0$ & $0$ \\
\hline
\hline
$c=1$ & $s=0$ & $\Sigma_c^{++}$   & $\ep(u^T C\gaf c)u$ & $1$ & $+1$ \\
      &       & $\Sigma_c^{+}$    & $\frac{1}{\sqrt{2}}\ep\{(u^T C\gaf c)d+(d^T C\gaf c)u\}$ & $1$ & $0$ \\
      &       & $\Sigma_c^{0}$    & $\ep(d^T C\gaf c)d$ & $1$ & $-1$ \\
\hline
      & $s=1$ & $\Xi_c^{\prime+}$ & $\frac{1}{\sqrt{2}}\ep\{(s^T C\gaf c)u+(u^T C\gaf c)s\}$ & $1/2$ & $+1/2$ \\
      &       & $\Xi_c^{\prime0}$ & $\frac{1}{\sqrt{2}}\ep\{(s^T C\gaf c)d+(d^T C\gaf c)s\}$ & $1/2$ & $-1/2$ \\
\hline
      & $s=2$ & $\Omega_c^{0}$    & $\ep(s^T C\gaf c)s$ & $0$ & $0$ \\
\hline
      & $s=0$ & $\Lambda_c^{+}$   & $\frac{1}{\sqrt{6}}\ep\{2(u^T C\gaf d)c+(u^T C\gaf c)d-(d^T C\gaf c)u\}$ & $0$ & $0$ \\
\hline
      & $s=1$ & $\Xi_c^{+}$       & $\frac{1}{\sqrt{6}}\ep\{2(s^T C\gaf u)c+(s^T C\gaf c)u-(u^T C\gaf c)s\}$ & $1/2$ & $+1/2$ \\
      &       & $\Xi_c^{0}$       & $\frac{1}{\sqrt{6}}\ep\{2(s^T C\gaf d)c+(s^T C\gaf c)d-(d^T C\gaf c)s\}$ & $1/2$ & $-1/2$ \\
\hline
\hline
$c=2$ & $s=0$ & $\Xi_{cc}^{++}$   & $\ep(c^T C\gaf u)c$ & $1/2$ & $+1/2$ \\
      &       & $\Xi_{cc}^{+}$    & $\ep(c^T C\gaf d)c$ & $1/2$ & $-1/2$ \\
\hline
      & $s=1$ & $\Omega_{cc}^{+}$ & $\ep(c^T C\gaf s)c$ & $0$ & $0$ \\
\hline
\end{tabular}
\caption{\label{tab:oct}\sl
Interpolating fields of SU(4) $20'$-plet (``octet-type'') baryons with spin
1/2. Throughout the color indices are suppressed, i.e.\ $\ep(x^T C\gaf y)z$ is
to be read as $\ep_{abc}(x_a^T C\gaf y_b)z_c$.}
\end{table}

The $20'$-plet decomposes into the standard $c=0$ ground floor which
transforms as an $8$ under SU(3) [lines 1-8 in Tab.\,\ref{tab:oct}],
the $c=1$ first floor which decomposes into a $6$ [lines 9-14] and a $\bar{3}$
[lines 15-17], and the $c=2$ second floor which transforms as a $3$ under
SU(3) [lines 18-20].
Regarding the $c=1$ level, it is worth noticing that the states of the $6$ are
symmetric under interchange of the two non-charmed quarks, whereas the states
of the $\bar{3}$ are antisymmetric under this interchange.
Here we adopt the rule that in the $20'$-plet the diquark $(x^T C\gaf y)$ is
antisymmetric under the interchange $x\leftrightarrow y$.

\begin{table}[!tb]
\centering
\begin{tabular}{|cc|cccc|}
\hline
charm & strange & baryon & interpolating field & $I$ & $I_z$ \\
\hline
\hline
$c=0$ & $s=0$ & $\Delta^{++}$          & $\ep(u^T C\gam u)u$ & $3/2$ & $+3/2$ \\
      &       & $\Delta^{+}$           & $\frac{1}{\sqrt{3}}\ep\{2(u^T C\gam d)u+(u^T C\gam u)d\}$ & $3/2$ & $+1/2$ \\
      &       & $\Delta^{0}$           & $\frac{1}{\sqrt{3}}\ep\{2(d^T C\gam u)d+(d^T C\gam d)u\}$ & $3/2$ & $-1/2$ \\
      &       & $\Delta^{-}$           & $\ep(d^T C\gam d)d$ & $3/2$ & $-3/2$ \\
\hline
      & $s=1$ & $\Sigma^{\star+}$      & $\frac{1}{\sqrt{3}}\ep\{2(u^T C\gam s)u+u^T C\gam u)s\}$ & $1$ & $+1$ \\
      &       & $\Sigma^{\star0}$      & $\frac{1}{\sqrt{3}}\ep\{(u^T C\gam d)s+(d^T C\gam s)u+(s^T C\gam u)d\}$ & $1$ & $0$ \\
      &       & $\Sigma^{\star-}$      & $\frac{1}{\sqrt{3}}\ep\{2(d^T C\gam s)d+d^T C\gam d)s\}$ & $1$ & $-1$ \\
\hline
      & $s=2$ & $\Xi^{\star0}$         & $\frac{1}{\sqrt{3}}\ep\{2(s^T C\gam u)s+(s^T C\gam s)u\}$ & $1/2$ & $+1/2$ \\
      &       & $\Xi^{\star-}$         & $\frac{1}{\sqrt{3}}\ep\{2(s^T C\gam d)s+(s^T C\gam s)d\}$ & $1/2$ & $-1/2$ \\
\hline
      & $s=3$ & $\Omega^{-}$           & $\ep(s^T C\gam s)s$ & $0$ & $0$ \\
\hline
\hline
$c=1$ & $s=0$ & $\Sigma_c^{\star++}$   & $\frac{1}{\sqrt{3}}\ep\{2(u^T C\gam c)u+u^T C\gam u)c\}$ & $1$ & $+1$ \\
      &       & $\Sigma_c^{\star+}$    & $\frac{1}{\sqrt{3}}\ep\{(u^T C\gam d)c+(d^T C\gam c)u+(c^T C\gam u)d\}$ & $1$ & $0$ \\
      &       & $\Sigma_c^{\star0}$    & $\frac{1}{\sqrt{3}}\ep\{2(d^T C\gam c)d+d^T C\gam d)c\}$ & $1$ & $-1$ \\
\hline
      & $s=1$ & $\Xi_c^{\star+}$       & $\frac{1}{\sqrt{3}}\ep\{(u^T C\gam s)c+(s^T C\gam c)u+(c^T C\gam u)s\}$ & $1/2$ & $+1/2$ \\
      &       & $\Xi_c^{\star0}$       & $\frac{1}{\sqrt{3}}\ep\{(d^T C\gam s)c+(s^T C\gam c)d+(c^T C\gam d)s\}$ & $1/2$ & $-1/2$ \\
\hline
      & $s=2$ & $\Omega_c^{\star0}$    & $\frac{1}{\sqrt{3}}\ep\{2(s^T C\gam c)s+(s^T C\gam s)c\}$ & $0$ & $0$ \\
\hline
\hline
$c=2$ & $s=0$ & $\Xi_{cc}^{\star++}$   & $\frac{1}{\sqrt{3}}\ep\{2(c^T C\gam u)c+(c^T C\gam c)u\}$ & $1/2$ & $+1/2$ \\
      &       & $\Xi_{cc}^{\star+}$    & $\frac{1}{\sqrt{3}}\ep\{2(c^T C\gam d)c+(c^T C\gam c)d\}$ & $1/2$ & $-1/2$ \\
      & $s=1$ & $\Omega_{cc}^{\star+}$ & $\frac{1}{\sqrt{3}}\ep\{2(c^T C\gam s)c+(c^T C\gam c)s\}$ & $0$ & $0$ \\
\hline
\hline
$c=3$ & $s=0$ & $\Omega_{ccc}^{++}$    & $\ep(c^T C\gam c)c$ & $0$ & $0$ \\
\hline
\end{tabular}
\caption{\label{tab:dec}\sl
Interpolating fields of SU(4) $20$-plet (``decuplet-type'') baryons with spin
3/2. Throughout the color indices are suppressed, i.e.\ $\ep(x^T C\gam y)z$ is
to be read as $\ep_{abc}(x_a^T C\gam y_b)z_c$.}
\end{table}

The structure of the $20$-plet is somewhat simpler, since each fixed-$c$ floor
has a unique transformation pattern under SU(3).
It contains the standard $c=0$ ground floor which transforms as a $10$ under
SU(3) [lines 1-10 in Tab.\,\ref{tab:dec}], the $c=1$ first floor which
transforms as a $6$ [lines 11-16], the $c=2$ second floor which transforms as
a $3$ [lines 17-19], and the $c=3$ one-point summit of the pyramid [line 20].
Here we adopt the rule that in the $20$-plet the diquark $(x^T C\gam y)$ is
symmetric under the interchange $x\leftrightarrow y$.

\begin{table}[!tb]
\centering
\begin{tabular}{|cc|cccc|}
\hline
charm & strange & baryon & interpolating field & $I$ & $I_z$ \\
\hline
\hline
$c=0$ & $s=1$ & $\Lambda^{\prime0}$     & $\frac{1}{\sqrt{3}}\ep\{(u^T C\gaf d)s+(s^T C\gaf u)d+(d^T C\gaf s)u\}$ & $0$ & $0$ \\
\hline
\hline
$c=1$ & $s=0$ & $\Lambda_c^{\prime+}$   & $\frac{1}{\sqrt{3}}\ep\{(u^T C\gaf d)c+(c^T C\gaf u)d+(d^T C\gaf c)u\}$ & $0$ & $0$ \\
\hline
      & $s=1$ & $\Xi_c^{\prime\prime+}$ & $\frac{1}{\sqrt{3}}\ep\{(u^T C\gaf s)c+(c^T C\gaf u)s+(s^T C\gaf c)u\}$ & $1/2$ & $+1/2$ \\
      &       & $\Xi_c^{\prime\prime0}$ & $\frac{1}{\sqrt{3}}\ep\{(d^T C\gaf s)c+(c^T C\gaf d)s+(s^T C\gaf c)d\}$ & $1/2$ & $-1/2$ \\
\hline
\end{tabular}
\caption{\label{tab:new}\sl
Interpolating fields of SU(4) $\bar{4}$-plet (``new-type'') baryons.
Throughout the color indices are suppressed, i.e.\ $\ep(x^T C\gaf y)z$ is to
be read as $\ep_{abc}(x_a^T C\gaf y_b)z_c$.}
\end{table}

The $\bar{4}$-plet decomposes into a $c=0$ ground floor which is an
SU(3) singlet [line 1 in Tab.\,\ref{tab:new}], and a $c=1$ first floor which
transforms as a $\bar{3}$ [lines 2-4].
In the former case the construction is based on the requirement that
$\Sigma^0\propto(us)d+(ds)u$,
$\Lambda^0\equiv\Lambda_8^0\propto2(ud)s+(us)d-(ds)u$, and
$\Lambda^{\prime0}\equiv\Lambda_0^0\propto(ud)s+(su)d+(ds)u$ would be mutually
orthogonal.
In the latter case the interpolator is antisymmetric under the interchange of
the two non-charmed quarks, if we adopt the rule that the diquark
$(x^T C\gaf y)$ is antisymmetric under the interchange $x\leftrightarrow y$.

\clearpage




\begin{thebibliography}{99} \itemsep-2pt


\bibitem{Fodor:2012gf}
  Z.~Fodor and C.~Hoelbling,
  Rev.\ Mod.\ Phys.\  {\bf 84}, 449 (2012) [arXiv:1203.4789].

\bibitem{Sheikholeslami:1985ij}
  B.~Sheikholeslami and R.~Wohlert,
  Nucl.\ Phys.\ B {\bf 259}, 572 (1985).
\bibitem{Luscher:1996sc}
  M.~Luscher, S.~Sint, R.~Sommer and P.~Weisz,
  Nucl.\ Phys.\ B {\bf 478}, 365 (1996) [hep-lat/9605038].

\bibitem{DeGrand:1998jq}
  T.A.~DeGrand, A.~Hasenfratz and T.G.~Kovacs [MILC Collaboration],
  hep-lat/9807002.
\bibitem{Bernard:1999kc}
  C.W.~Bernard and T.~DeGrand,
  Nucl.\ Phys.\ Proc.\ Suppl.\ {\bf 83}, 845 (2000) [hep-lat/9909083].
\bibitem{Stephenson:1999ns}
  M.~Stephenson, C.~DeTar, T.A.~DeGrand and A.~Hasenfratz,
  Phys.\ Rev.\ D {\bf 63}, 034501 (2001) [hep-lat/9910023].
\bibitem{Zanotti:2001yb}
  J.M.~Zanotti {\it et al.} [CSSM Collab.],
  Phys.\ Rev.\ D {\bf 65}, 074507 (2002) [hep-lat/0110216].
\bibitem{DeGrand:2002vu}
  T.~DeGrand, A.~Hasenfratz and T.G.~Kovacs,
  Phys.\ Rev.\ D {\bf 67}, 054501 (2003) [hep-lat/0211006].

\bibitem{Capitani:2006ni}
  S.~Capitani, S.~Durr and C.~Hoelbling,
  JHEP {\bf 0611}, 028 (2006) [hep-lat/0607006].
\bibitem{Durr:2007cy}
  S.~Durr,
  Comput.\ Phys.\ Commun.\ {\bf 180}, 1338 (2009) [arXiv:0709.4110].

\bibitem{Durr:2010aw}
  S.~Durr, Z.~Fodor, C.~Hoelbling, S.~D.~Katz, S.~Krieg, T.~Kurth, L.~Lellouch
  and T.~Lippert {\it et al.},
  JHEP {\bf 1108}, 148 (2011) [arXiv:1011.2711].

\bibitem{Durr:2010ch}
  S.~Durr and G.~Koutsou,
  Phys.\ Rev.\ D {\bf 83}, 114512 (2011) [arXiv:1012.3615].
\bibitem{Bietenholz:1999km}
  W.~Bietenholz and I.~Hip,
  Nucl.\ Phys.\ B {\bf 570}, 423 (2000) [hep-lat/9902019].


\bibitem{Bietenholz:2010az}
  W.~Bietenholz, M.~Gockeler, R.~Horsley, Y.~Nakamura, D.~Pleiter,
  P.~E.~L.~Rakow, G.~Schierholz and J.~M.~Zanotti,
  Phys.\ Lett.\ B {\bf 687}, 410 (2010) [arXiv:1002.1696].
\bibitem{QCDSF:2011aa}
  G.~S.~Bali {\it et al.}  [QCDSF Collab.],
  Phys.\ Rev.\ Lett.\  {\bf 108}, 222001 (2012) [arXiv: 1112.3354].


\bibitem{Gusken:1989ad}
  S.~Gusken, U.~Low, K.~H.~Mutter, R.~Sommer, A.~Patel and K.~Schilling,
  Phys.\ Lett.\ B {\bf 227}, 266 (1989).

\bibitem{Colangelo:2010et}
  G.~Colangelo {\it et al.} [FLAG],
  Eur.\ Phys.\ J.\ C {\bf 71}, 1695 (2011) [arXiv:1011.4408].

\bibitem{Nakamura:2010zzi}
  K.~Nakamura {\it et al.}  [Particle Data Group Collab.],
  J.\ Phys.\ G G {\bf 37}, 075021 (2010).

\bibitem{DelDebbio:2005qa}
  L.~Del Debbio, L.~Giusti, M.~Luscher, R.~Petronzio and N.~Tantalo,
  JHEP {\bf 0602}, 011 (2006) [hep-lat/0512021].


\bibitem{Lee:1998cx}
  F.~X.~Lee and D.~B.~Leinweber,
  Nucl.\ Phys.\ Proc.\ Suppl.\  {\bf 73}, 258 (1999) [hep-lat/9809095].
\bibitem{Leinweber:2004it}
  D.~B.~Leinweber, W.~Melnitchouk, D.~G.~Richards, A.~G.~Williams and
  J.~M.~Zanotti,
  Lect.\ Notes Phys.\  {\bf 663}, 71 (2005) [nucl-th/0406032].

\bibitem{Bowler:1996ws}
  K.~C.~Bowler {\it et al.}  [UKQCD Collab.],
  Phys.\ Rev.\ D {\bf 54}, 3619 (1996) [hep-lat/9601022].
\bibitem{Edwards:2011jj}
  R.~G.~Edwards, J.~J.~Dudek, D.~G.~Richards and S.~J.~Wallace,
  Phys.\ Rev.\ D {\bf 84}, 074508 (2011) [arXiv:1104.5152].



\bibitem{Lewis:2001iz}
  R.~Lewis, N.~Mathur and R.~M.~Woloshyn,
  Phys.\ Rev.\ D {\bf 64}, 094509 (2001) [hep-ph/0107037].

\bibitem{Mathur:2002ce}
  N.~Mathur, R.~Lewis and R.~M.~Woloshyn,
  Phys.\ Rev.\ D {\bf 66}, 014502 (2002) [hep-ph/0203253].

\bibitem{Flynn:2003vz}
  J.~M.~Flynn {\it et al.} [UKQCD Collab.],
  JHEP {\bf 0307}, 066 (2003) [hep-lat/0307025].

\bibitem{Na:2008hz}
  H.~Na and S.~Gottlieb,
  PoS LATTICE {\bf 2008}, 119 (2008) [arXiv:0812.1235].

\bibitem{Liu:2009jc}
  L.~Liu, H.~-W.~Lin, K.~Orginos and A.~Walker-Loud,
  Phys.\ Rev.\ D {\bf 81}, 094505 (2010) [arXiv:0909.3294].

\bibitem{Lin:2010wb}
  H.~-W.~Lin, S.~D.~Cohen, L.~Liu, N.~Mathur, K.~Orginos and A.~Walker-Loud,
  Comput.\ Phys.\ Commun.\  {\bf 182}, 24 (2011) [arXiv:1002.4710].

\bibitem{Mohler:2011ke}
  D.~Mohler and R.~M.~Woloshyn,
  Phys.\ Rev.\ D {\bf 84}, 054505 (2011) [arXiv:1103.5506].

\bibitem{Namekawa:2011wt}
  Y.~Namekawa {\it et al.}  [PACS-CS Collab.],
  Phys.\ Rev.\ D {\bf 84}, 074505 (2011) [arXiv:1104.4600].

\bibitem{Lin:2011ti}
  H.~-W.~Lin,
  Chin.\ J.\ Phys.\  {\bf 49}, 827 (2011) [arXiv:1106.1608].

\bibitem{Alexandrou:2012xk}
  C.~Alexandrou, J.~Carbonell, D.~Christaras, V.~Drach, M.~Gravina and
  M.~Papinutto,
  arXiv:1205.6856 [hep-lat].

\bibitem{Briceno:2012wt}
  R.~A.~Briceno, H.~-W.~Lin and D.~R.~Bolton,
  Phys.\ Rev.\ D {\bf 86}, 094504 (2012) [arXiv:1207.3536].


\bibitem{ElKhadra:1996mp}
  A.~X.~El-Khadra, A.~S.~Kronfeld and P.~B.~Mackenzie,
  Phys.\ Rev.\ D {\bf 55}, 3933 (1997) [hep-lat/9604004].

\bibitem{Hasenfratz:1993sp}
  P.~Hasenfratz and F.~Niedermayer,
  Nucl.\ Phys.\ B {\bf 414}, 785 (1994) [hep-lat/9308004].
\bibitem{Bietenholz:1995cy}
  W.~Bietenholz and U.~J.~Wiese,
  Nucl.\ Phys.\ B {\bf 464}, 319 (1996) [hep-lat/9510026].
\bibitem{Hasenfratz:2000xz}
  P.~Hasenfratz, S.~Hauswirth, K.~Holland, T.~Jorg, F.~Niedermayer and U.~Wenger,
  Int.\ J.\ Mod.\ Phys.\ C {\bf 12}, 691 (2001) [hep-lat/0003013].
\bibitem{Hasenfratz:2002rp}
  P.~Hasenfratz, S.~Hauswirth, T.~Jorg, F.~Niedermayer and K.~Holland,
  Nucl.\ Phys.\  B {\bf 643}, 280 (2002) [hep-lat/0205010].

\bibitem{Golterman:1984cy}
  M.~F.~L.~Golterman and J.~Smit,
  Nucl.\ Phys.\ B {\bf 245}, 61 (1984).
\bibitem{Adams:2010gx}
  D.~H.~Adams,
  Phys.\ Lett.\ B {\bf 699}, 394 (2011) [arXiv:1008.2833].
\bibitem{Hoelbling:2010jw}
  C.~Hoelbling,
  Phys.\ Lett.\ B {\bf 696}, 422 (2011) [arXiv:1009.5362].
\bibitem{Creutz:2010bm}
  M.~Creutz, T.~Kimura and T.~Misumi,
  JHEP {\bf 1012}, 041 (2010) [arXiv:1011.0761].
\bibitem{deForcrand:2012bm}
  P.~de Forcrand, A.~Kurkela and M.~Panero,
  JHEP {\bf 1204}, 142 (2012) [arXiv:1202.1867].

\bibitem{Durr:2011ed}
  S.~Durr and G.~Koutsou,
  Phys.\ Rev.\ Lett.\  {\bf 108}, 122003 (2012) [arXiv:1108.1650].

\end{thebibliography}
\end{document}